\documentclass[twocolumn, aps, pra, superscriptaddress, 10pt]{revtex4-1}
\usepackage{amsmath,amssymb,mathtools,xcolor,bbold}
\usepackage{graphicx}
\usepackage{hyperref}
\usepackage[T1]{fontenc}
\usepackage[utf8]{inputenc}
\usepackage{enumitem}
\usepackage{placeins}
\usepackage{color} 
\usepackage[abs]{overpic}

\graphicspath{ {./figures/} }

\begin{document}
\title{Persistent currents by reservoir engineering}

\author{Maximilian Keck}
\affiliation{NEST, Scuola Normale Superiore and Istituto Nanoscienze-CNR, I-56126 Pisa, Italy}

\author{Davide Rossini}
\affiliation{Dipartimento di Fisica, Universit\`a di Pisa and INFN, Largo Pontecorvo 3, I-56127 Pisa, Italy}

\author{Rosario Fazio}
\affiliation{ICTP, Strada Costiera 11, I-34151 Trieste, Italy}
\affiliation{NEST, Scuola Normale Superiore and Istituto Nanoscienze-CNR, I-56126 Pisa, Italy}

\begin{abstract}
  We demonstrate that persistent currents can be induced in a quantum system in contact with
  a structured reservoir, without the need of any applied gauge field.
  The working principle of the mechanism leading to their presence is based on the extension
  to the many-body scenario of non-reciprocal Lindblad dynamics,
  recently put forward by Metelmann and Clerk in Phys. Rev. X {\bf 5}, 021025 (2015):
  Non-reciprocity can be generated by suitably balancing coherent interactions with
  their corresponding dissipative version, induced by the coupling to a common
  structured environment, such to make total interactions directional.
  Specifically, we consider an interacting spin/boson model in a ring-shaped one-dimensional
  lattice coupled to an external bath. By employing a combination of cluster mean-field,
  exact diagonalization and matrix-product-operator techniques, we show that solely dissipative
  effects suffice to engineer steady states with a persistent current that survives
  in the limit of large systems.
  We also verify the robustness of such current in the presence of additional dissipative
  or Hamiltonian perturbation terms.
\end{abstract}

\maketitle

\section{Introduction}
\label{sec:introduction}

The eigenstates of a quantum particle in a ring pierced by a magnetic field support a non-zero
circulating current which is periodic in the external magnetic flux. The period of the oscillation
is the flux quantum $hc/q$, defined only through the Plank's constant, the speed of light
and the charge $q$ of the particle. Since its discovery~\cite{Yang_1967}, this cornerstone result
of quantum mechanics has constantly pervaded, in various forms, many different areas of physics.
Periodic oscillations in the critical temperature of hollow superconducting cylinders have been
observed, standing as a clear signature of macroscopic quantum coherence~\cite{Tinkham_1996}.
Persistent currents circulating in small one-dimensional (1D) rings have been one of the most attractive
phenomena in mesoscopics, since they are a direct manifestation of quantum coherence~\cite{Imry_1997}.
Such type of phenomenology has been more recently extended to the realm of cold-atom physics,
thanks to the experimental and theoretical advances in the study and manipulation of currents
induced in ring traps by the application of a rotating barrier~\cite{Phillips}, or artificial 
gauge fields~\cite{Dalibard_2011, Spielman_2014}.
This led to the growth of a research field called atomtronics, in which optical circuits
of different spatial shape and intensity have been devised to exploit interesting analogies
between electronics (and its applications) and confined
atoms~\cite{Amico_2005, Seaman_2007, Amico_2014, Aghamalyan_2015, Dumke_2016, Haug_2018}.
A net advantage of such more recent alternative resides in the extremely high degree of tunability,
controllability and readability of the microscopic details at the level of a single atom,
together with the possibility to keep coherence under control for macroscopic time scales 
up to the order of few seconds.
Several of these ingredients are important for the problem we are going to introduce in the present work.

In the cases discussed so far, the presence of circulating currents require an external gauge field 
and the establishing of a substantial degree of quantum coherence over the whole ring.
The purpose of this paper is to show that it is possible to generate currents in rings even by reservoir 
engineering. Neither a real/synthetic magnetic field nor quantum coherence are required. 

Recent years have seen the birth of a new quantum technology era, accompanied by the flourishing
of a wealth of different possibilities, which enable to engineer various states of matter in several
intriguing conditions. One of them deals with the investigation of driven-dissipative
many-body systems~\cite{Kasprak_2006, Esslinger_2010, Carusotto_2013, Fitzpatrick_2017}. Available platforms include
coupled QED cavities~\cite{Hartmann_2008, TomadinRev_2010, Houck_2012, Noh_2016}
or optomechanical arrays~\cite{Zhou_2008, Marquardt_2013, Zhu_2013}, as well as atomic and molecular optical systems
such as Rydberg atoms~\cite{Saffmann_2010} or trapped ions~\cite{Muller_2012}.
Dissipation however is not necessarily detrimental. Indeed, motivated by the seminal works
of Diehl {\em et al.}~\cite{Diehl_2008} and Verstraete {\em et al.}~\cite{Verstraete_2009},
a constantly growing theoretical activity started to carefully scrutinise the possibility to prepare 
complex many-body quantum states through the engineering of a system-bath coupling.
This could lead, for example, to the emergence of novel states of quantum matter or of intriguing
topological features~\cite{Diehl_2011, Bardyn_2013, Iemini_2015}.
In the field of quantum optics, the employment of a synthetic dimension~\cite{Ozawa_2016}
and the study of chiral phenomena (i.e., chiral quantum optics) have recently gained
considerable traction~\cite{Pichler_2015, Lodahl_2017, Vermersch_2017}.

In the same spirit as in the dissipative preparation of many-body states, here we show how to realize
a steady-state many-body current-carrying state. 
Our approach stems from a recent proposal by Metelmann and Clerk~\cite{Metelmann_2015},
where a general method to break reciprocality was developed and studied for the case of two cavities coupled to each other.
Ref.~\cite{Metelmann_2015} sets up the method to realize non-reciprocal photon transmission and amplification
by matching coherent and dissipative parts of the dynamics. We adopt the same approach and extend it to many-body systems.
Differently from current states induced by gauge fields, the ones obtained by reservoir engineering do not vanish
in the thermodynamic limit. This is probably the most striking difference, showing that dissipative realization of 
persistent current is profoundly different from what we are used to observe.  

The paper is organized as follows. In Sec.~\ref{sec:model} we first introduce the model of interacting bosons
that we are going to analyse and its mapping to a spin-$1/2$ system. We discuss in details the properties of the reservoir,
we then derive the expression for particle current from the equation of motion of the magnetization and introduce
the various quantities that will be analysed later. In Sec.~\ref{sec:flux_based_currents} we briefly address the emergence 
of currents induced by a magnetic flux in a ring. This Section is important for a comparison with the dissipative mechanism.
The non-reciprocity induced current in our driven-dissipative model is addressed in Sec.~\ref{sec:CurrRes},
both at the mean-field level and numerically.
We first adopt a cluster mean-field (CMF) model, as it is the simplest case that can sustain such a current,
and within this approach we study the features of the emerging current, discussing both the possibility
to explain the phenomenon as well as the limits of this approach.
Later, after supporting the CMF solution with numerical exact diagonalization and matrix product operators (MPOs),
we study the robustness of the proposed current if the minimal model is perturbed,
showing that this phenomenon withstands imperfections.
Finally, Sec.~\ref{sec:conclusion} is devoted to a summary of our results.

\section{Model}
\label{sec:model}

A good candidate that is amenable to test our proposal is a lattice of coupled QED cavities. These indeed represent
the ideal platform in which it is possible to manipulate the various internal degrees of freedom and the
coupling with ancillary systems, thus enabling reservoir engineering.
In its simplest configuration, the system consists of $L$ cavities coupled in a ring-shaped fashion,
whose physics is well captured by the following 1D lattice Hamiltonian, {written
in the rotating frame}~\cite{Hartmann_2008, TomadinRev_2010}:
\begin{equation}
  H = \sum_j \big( J_j d^{\dagger}_{j} d_{j+1} + \lambda_j d^{\dagger}_{j}d^{\dagger}_{j+1} + \textrm{h.c.} \big)
  + U_j n_j (n_j - 1) .
  \label{eq:HamModel}
\end{equation}

The quadratic part in the first sum denotes nearest-neighbor tunneling terms,
which naturally take into account the coupling between adjacent cavities
through tunable site-dependent complex parameters $J_j$ and $\lambda_j$,
while the second sum stands for a local repulsive interaction of strength $U_j>0, \; \forall j$.
The operators $d^{(\dagger)}_j$ create/destroy a boson on a given cavity $j$
($j=1,\ldots,L$) and satisfy the usual bosonic commutation relations,
while $n_j = d^\dagger_j d_j$ is the corresponding onsite number operator.

\begin{figure}[!t]
  \includegraphics[width=\columnwidth]{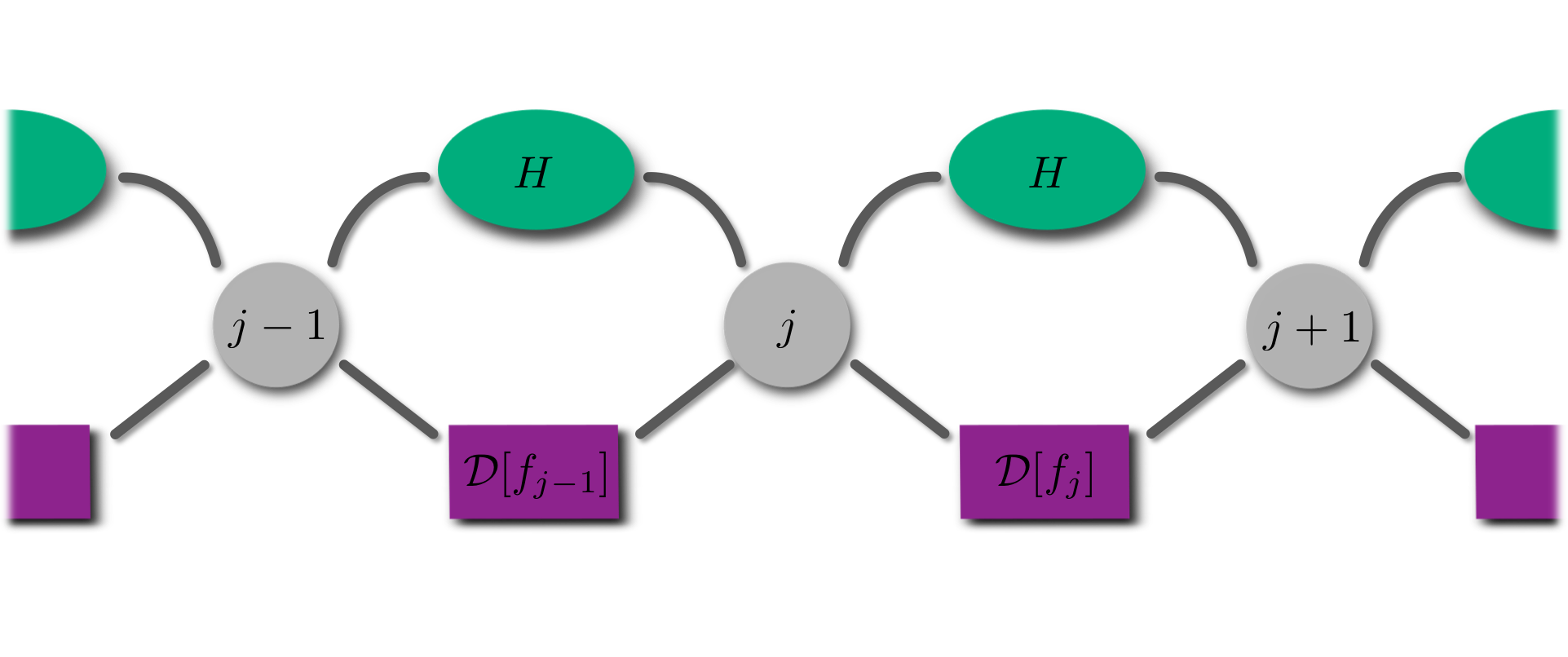}
  \caption{A sketch illustrating our setup. The local degrees of freedom (on sites $j = 1,2, \dots, L$)
    have the possibility to interact both unitarily, with a nearest neighbor interaction induced by the Hamiltonian $H$,
    and dissipatively, via an engineered reservoir induced by the Markovian dissipator $\mathcal{D}[f_j(\{d\})]$.}
  \label{fig:sketch}
\end{figure}

For this kind of quantum optical setup, the effect of an external environment can be faithfully modeled as Markovian,
and is well described by a master equation in the Lindblad form~\cite{Petruccione_2002}:
\begin{equation}
  \dfrac{\mathrm{d}}{\mathrm{d}t} \rho = - i [H,\rho]
  + \sum_{j} \kappa_{j} \, \mathcal{D}[d_{j}] \, \rho + \sum_{j} \mathcal{D}[f_{j}(\{d\})] \, \rho ,
  \label{eq:MasterModel}
\end{equation}
where the term
\begin{equation}
  \mathcal{D}[O]\rho \equiv O\rho O^{\dagger} - \tfrac{1}{2}\{O^{\dagger}O,\rho\}
  \label{eq:Diss_def}
\end{equation}
encodes the dissipative part of the Liouvillian superoperator.
Hereafter we adopt units of $\hbar = 1$, denote with $[\cdot,\cdot]$ the commutator
and with $\{ \cdot, \cdot \}$ the anticommutator.
The first term in the r.h.s.~of Eq.~\eqref{eq:MasterModel} describes the coherent evolution
driven by the Hamiltonian~\eqref{eq:HamModel}.
The second term is typically due to the presence of the (unstructured) environment, and 
corresponds to a local (on-site) coupling to an independent bath
of strength $\kappa_j>0$, whose effect is to incoherently deplete the corresponding cavity (or site).
The third term accounts for an engineered two-site dissipation, since
\begin{equation}
 f_{j}(\{d\}) = \alpha_{j} d_{j} + \beta_{j} d^{\dagger}_{j} + \gamma_{j} d_{j+1} + \delta_{j} d^{\dagger}_{j+1}
  \label{eq:CorrBath}
\end{equation}
describes a correlated environment which, as we shall see below, will give rise to a persistent steady-state current.
{This term effectively describes the interaction of two sites ($j$ and $j+1$) with an engineered reservoir;
  this can be represented, e.g., by another auxiliary cavity mode (say, $c^{(\dagger)}_{j,j+1}$) that is coupled
  with the two principal modes ($d^{(\dagger)}_j$ and $d^{(\dagger)}_{j+1}$) via a quadratic Hamiltonian~\cite{Metelmann_2015}.
  Such quadratic interaction can be implemented via standard quantum optical techniques.
  Tracing out the auxiliary system, one recovers the form in the third term of the master equation~\eqref{eq:MasterModel}.}
A sketch the system described by Eq.~\eqref{eq:MasterModel}, reduced to its essentials, is shown in Fig.~\ref{fig:sketch}.

The model defined by Eqs.~\eqref{eq:HamModel}-\eqref{eq:Diss_def} is a direct generalization
to many sites of the one presented in Ref.~\cite{Metelmann_2015} for a pair of coupled cavities,
which was shown to exhibit directionality after a suitable tuning of the various coupling parameters
with the correlated bath [namely, the $\alpha_j$, $\beta_j$, $\gamma_j$, $\delta_j$ appearing
in Eq.~\eqref{eq:CorrBath}].
As we shall see below, an analogous mechanism can be established in our case,
in such a way that the resulting equation of motion for $d_{j}$ does not depend on $d_{j+1}$,
while that for $d_{j+1}$ depends on $d_{j}$ (for further details we refer to App.~\ref{app:BH model}).
The additional local interaction term $U_{j} n_{j} (n_{j} - 1)$ has no influence on this. 

In the Heisenberg picture, the dynamics of the expectation value
of a generic onsite operator $d_j$ takes the form
\begin{align}
  \dfrac{\mathrm{d}}{\mathrm{d}t} d_j
  &= d_j [ -2iU - \Pi ] \nonumber \\
  &+ d_{j-1} (-iJ^* + \eta^*/2) + d^{\dagger}_{j-1} (-i\lambda - \xi/2) \nonumber \\
  &+d_{j+1} (-iJ + \eta/2) + d^{\dagger}_{j+1} (-i\lambda + \xi/2),
  \label{eq:EoM_BH}
\end{align}
where
\begin{eqnarray}
  \Pi & = & (\kappa + |\alpha|^2 + |\beta|^2 + |\gamma|^2 + |\delta|^2)/2, \\
  \eta & = & \alpha\gamma^*-\delta\beta^*, \\
  \xi & = & \alpha\delta^{*}-\gamma^{*}\beta.
\end{eqnarray}
Choosing for example $\eta = 2iJ$ and $\xi = 2i \lambda$, we obtain
\begin{align}
  \dfrac{\mathrm{d}}{\mathrm{d}t} d_j
  &= d_j \big[ -2iU - \Pi \big] \nonumber \\
  &+ d_{j-1} (-2iJ^*) + d^{\dagger}_{j-1} (-2i\lambda),
  \label{eq:EoM_BH_unidirectional}
\end{align}
which elucidates the fact that it is possible to have situations where site $j$ is influenced by the previous
site $j-1$, but not by site $j+1$ (or vice-versa), thus breaking reciprocality of the system~\cite{Metelmann_2015}.
Notice that, here and in the rest of the paper, the various parameters entering the Hamiltonian~\eqref{eq:HamModel}
and the master equation~\eqref{eq:MasterModel} have been taken independent of the site index $j$.
This follows from having supposed that all the cavities are equal and enforcing translational invariance,
an assumption which does not spoil generality in our purposes.

In the limit of very strong repulsion, such that the energy scale
fixed by the onsite interaction $U$ is much larger than all the other ones in the system,
only the two levels with zero ($|0\rangle_j$) and one ($|1\rangle_j$) boson
are relevant for each bosonic mode.
The model can be thus mapped onto an effective spin-$1/2$ system.
Specifically, by setting $d^\dagger_j \to \sigma^+_j$ (and $d_j \to \sigma^-_j$),
it is immediate to see that the Hamiltonian in Eq.~\eqref{eq:HamModel} takes the form
\begin{equation}
  \label{eq:hamiltonian}
  \tilde H = \sum_{j} \big( J\sigma^{+}_{j}\sigma^{-}_{j+1} + \lambda \sigma^{+}_{j}\sigma^{+}_{j+1} + \textrm{h.c.} \big)
  + \mu \sigma^{+}_{j}\sigma^{-}_{j},
\end{equation}
where $\sigma^\alpha_j$ ($\alpha=x,y,z$) are the $2 \times 2$ Pauli matrices
on site $j$, and $\sigma^\pm_j = \tfrac12 \big( \sigma^x_j \pm i \sigma^y_j \big)$
are the corresponding raising/lowering spin operators.
Notice that in Eq.~\eqref{eq:hamiltonian} we have added a chemical potential term in $\mu$,
which replaces the energy offset introduced by the local repulsion term in $U$.
In the spin-$1/2$ limit, the full master equation~\eqref{eq:MasterModel} thus becomes
\begin{equation}
  \dfrac{\mathrm{d}}{\mathrm{d}t} \rho = - i [\tilde H,\rho]
  + \kappa \sum_{j} \mathcal{D}[\sigma^{-}_{j}] \, \rho + \sum_{j} \mathcal{D}[\tilde{f}_j(\{\sigma\})] \, \rho ,
\end{equation}
where
\begin{equation}
  \tilde{f}_j(\{\sigma\}) = \alpha \sigma^{-}_{j} + \beta \sigma^{+}_{j} + \gamma \sigma^{-}_{j+1} + \delta \sigma^{+}_{j+1}
\end{equation}
denotes the Lindblad jump operator associated to the correlated bath.

\subsection{Particle current}

The main subject of our analysis is not the directionality of the equations of motions themselves,
but rather the possibility to measure macroscopic effects in the system induced
by directionality in the dissipation, e.g. a current.
For this purpose, we concentrate on the onsite magnetization of the effective spin-$1/2$ model.
Specializing Eq.~\eqref{eq:EoM_BH} to the operator $\sigma^z_j$ and applying With the commutation relations
of the Pauli matrices
$[\sigma^{\alpha}_j,\sigma^{\beta}_\ell]= 2i \, \delta_{j\ell} \, \varepsilon_{\alpha\beta\gamma} \, \sigma^{\gamma}_j$,
where $\varepsilon_{\alpha\beta\gamma}$ is the Levi-Civita symbol, the following equation of motion can be easily obtained:
\begin{eqnarray}
  \dfrac{\mathrm{d}}{\mathrm{d}t} \sigma^{z}_{j}  & = & \nabla_j {\cal I}^{J}_{j} + \nabla_j {\cal I}^\xi_j
  - \big( {\cal I}^{\eta}_{j} + {\cal I}^{\eta}_{j+1} \big) + \big( {\cal I}^{\lambda}_{j} + {\cal I}^{\lambda}_{j+1} \big) \nonumber \\
  &&  - \big( \sigma^{z}_{j}+1 \big) (\Gamma^{-} + \kappa) - \big( \sigma^{z}_{j}-1 \big) \Gamma^{+}.
  \label{eq:MagCur}
\end{eqnarray}

Before going on with our analysis, it is useful to comment on the various terms
entering the r.h.s.~of Eq.~\eqref{eq:MagCur}.
The first and the third contributions,
$\nabla_j {\cal I}^J_j \equiv {\cal I}^{J}_{j} - {\cal I}^{J}_{j+1}$
and $\big( {\cal I}^{\eta}_{j} + {\cal I}^{\eta}_{j+1} \big)$,
express two forms of circulating currents
\begin{equation}
  {\cal I}^{J}_{j} \equiv 2i \big( J\sigma^{-}_{j}\sigma^{+}_{j-1} - \textrm{h.c.} \big) , \quad
  {\cal I}^{\eta}_{j} \equiv \big( \! -\!\eta \, \sigma^{+}_{j}\sigma^{-}_{j-1} +\textrm{h.c.} \big) ,
\end{equation}
induced by the tunneling term $J$ in the unitary dynamics
and by a dissipative contribution $\eta = \alpha\gamma^{*}-\beta^{*}\delta$, respectively.
On the other hand, the fourth and the second contributions,
$\big( {\cal I}^{\lambda}_{j} + {\cal I}^{\lambda}_{j+1} \big)$,
and $\nabla_j {\cal I}^\xi_j \equiv {\cal I}^\xi_j - {\cal I}^\xi_{j+1}$,
with
\begin{equation}
  {\cal I}^{\lambda}_{j} \equiv 2i \big( \lambda\sigma^{+}_{j}\sigma^{+}_{j-1} - \textrm{h.c.} \big) , \quad
  {\cal I}^{\xi}_{j} \equiv \big( \xi\sigma^{-}_{j}\sigma^{-}_{j-1} + \textrm{h.c.} \big),
\end{equation}
are associated to the pairing term $\lambda$ in the unitary dynamics
and to another dissipative contribution $\xi =\alpha\delta^{*}-\gamma^{*}\beta$,
and are not inducing a circulating current.
The last two terms of Eq.~\eqref{eq:MagCur} admit an easy interpretation:
they describe the on-site dissipation driving the system locally into one of the eigenstates of $\sigma^z$
with eigenvalue $+1$ or $-1$, respectively with strength
\begin{eqnarray}
  \Gamma^+ & = & |\beta|^2 + |\delta|^2,\\
  \Gamma^- + \kappa & = & |\alpha|^2 + |\gamma|^2 + \kappa.
\end{eqnarray}

If is useful to stress that, in this context, by ``circulating currents'' we mean those currents
that create an excitation on one site simultaneously with destroying an excitation at a neighboring site,
thus resulting in a flow of excitations.
Conversely, by ``non-circulating currents'' we mean those currents that create or destroy excitations
at two neighboring sites simultaneously and therefore not resulting in a flow of excitations.
Furthermore, it is also meaningful to differentiate between ``conserved currents'',
i.e., those that fulfill a continuity equation, which, on the lattice, takes the form
\begin{equation}
  \langle \nabla_j {\cal I}_j \rangle = \langle {\cal I}_j \rangle - \langle {\cal I}_{j+1} \rangle
  = \dfrac{\mathrm{d}}{\mathrm{d}t} \langle n_j \rangle,
\end{equation}
and those that do not fulfill such an equation (the ``non-conserved currents'').
Notice that the averages in square brackets denote expectation values over the steady-state (SS)
density matrix $\rho_{\rm SS} = \lim_{t \to \infty} \rho(t)$, that is, $\langle A \rangle = {\rm Tr}[A \, \rho_{\rm SS}]$.
The way Eq.~\eqref{eq:MagCur} is written makes it clear that currents induced by $J$ and $\eta$ are
conserved ones, while the other two contributions (induced by $\lambda$ and $\xi$) are not.

\begin{figure*}[!t]
    \includegraphics[width=0.68\columnwidth]{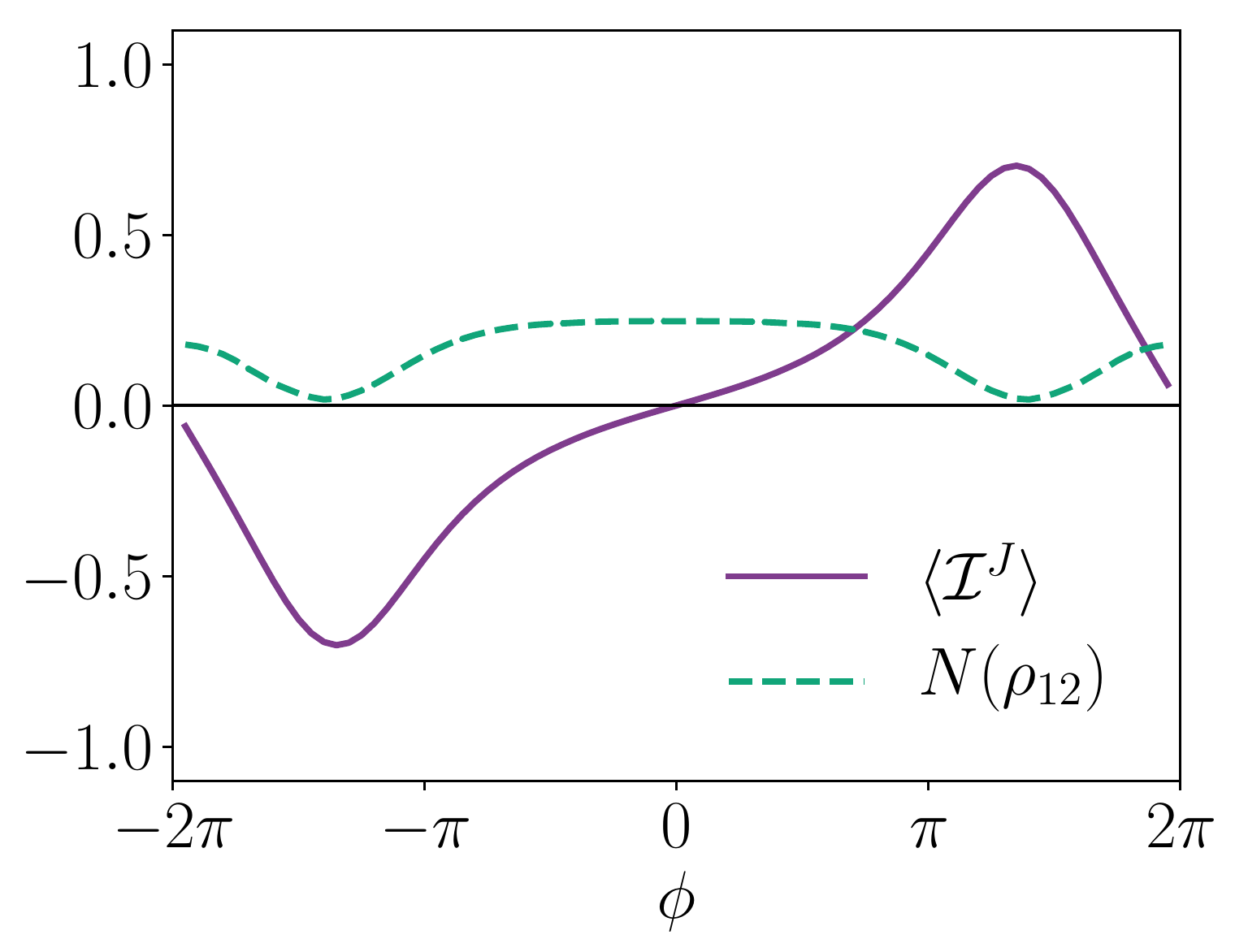}
    \includegraphics[width=0.68\columnwidth]{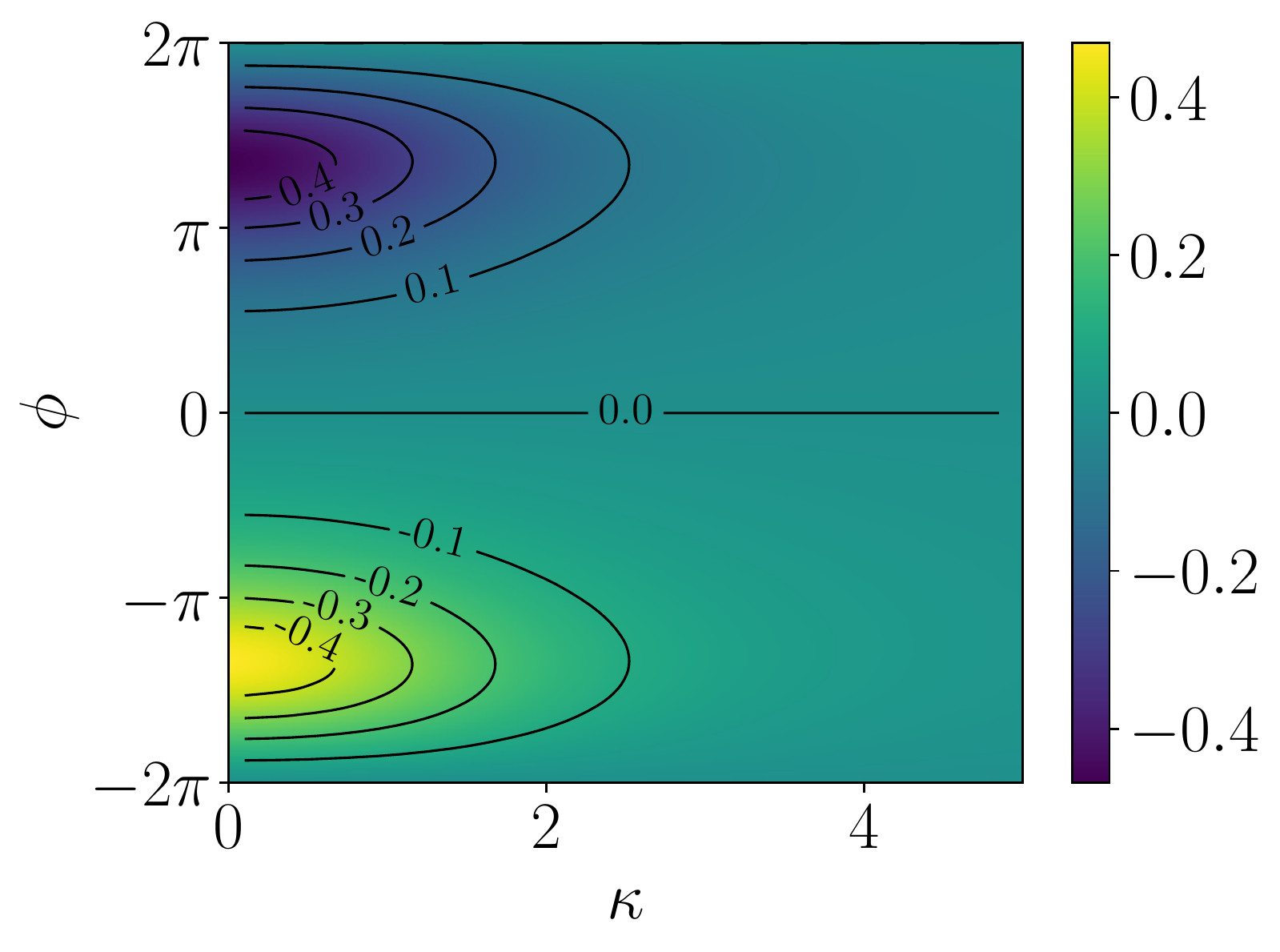}
    \includegraphics[width=0.68\columnwidth]{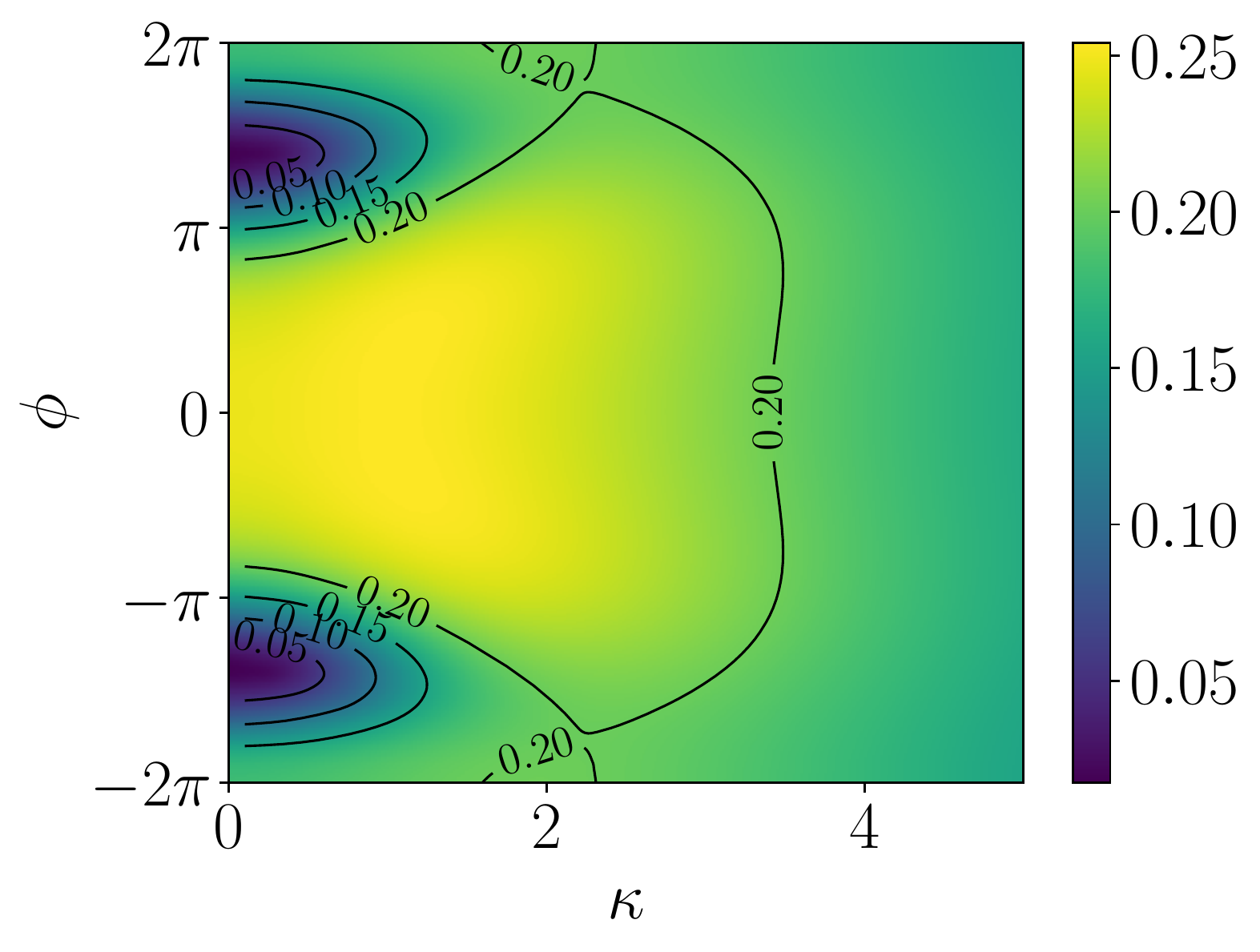}
    \caption{Flux-induced particle current in a chain of $L=4$ spins, described by
      the Hamiltionian~\eqref{eq:hamiltonian} with $J=1.5$ and $\lambda=0.5$.
      Left panel: steady-state current $\langle \cal{I}^J \rangle$ {(dark magenta, continuous line)}
      and negativity $N(\rho_{12})$ {(light green, dashed line)}
      induced by the flux $\phi$, in the presence of on-site dissipation of strength $\kappa=0.1$.
      Middle and right Panels: steady-state current (middle) and negativity (right panel),
      plotted as a function of both dissipation strength $\kappa$ and flux $\phi$.
      The energy scale is fixed by setting $\mu=1$ (this choice will be adopted in the rest of the paper).
      Notice that, due to translational invariance, the measured current is independent of the position $j$,
      therefore the corresponding index has been omitted.}
      \label{fig:steady_state_local_dissipation_phi_kappa}
\end{figure*}

\subsection{Quantum correlations}

Another property we are going to address is the amount of quantum correlations~\cite{Nielsen-Chuang}
that characterize the system, which will enable us to understand
some interesting features and differences between the emerging currents.
For pure states $|\psi\rangle$ the von Neumann entropy $S(\rho) = - \rm{Tr} \big[ \rho \ln \rho \big]$
of the reduced density matrix $\rho_{A} = {\rm Tr}_B \big[ |\psi\rangle \langle \psi| \big]$
of a generic bipartition $A|B$ gives an operative prescription for a good entanglement measure.
Unfortunately for mixed states, as the steady state $\rho_{\rm SS}$ of a system coupled to some environment,
it is not generally possible to define a good bipartite entanglement measure
through a simple closed formula. 

We will thus adopt different indicators of quantum correlations on $\rho_{\rm SS}$.
One of them is the negativity~\cite{Peres-Horodecki}
\begin{equation}
  N(\rho) = \tfrac 12 \big( \| \rho^{\Gamma_{\rm{A}}} \| - 1 \big) ,
\end{equation}
where $\| \cdot \|$ is the trace norm, and 
$\rho^{\Gamma_{\rm{A}}}$ denotes the partial transpose of $\rho$ with respect to subsystem $\rm{A}$.
While states with non-zero negativity are entangled, the converse is not always true;
thus the measure of $N(\rho)$ only provides a sufficient criterion for entanglement detection.
We will also analyze the purity~\cite{Nielsen-Chuang}
\begin{equation}
  P(\rho) = \rm{Tr} \big[ \rho^2 \big] ,
\end{equation}
giving information on the mixedness of a quantum state, being equal to $1$ for pure states,
and to $1/d$ for completely mixed states in a Hilbert space of dimension $d$.

\section{Currents induced by gauge fields}
\label{sec:flux_based_currents}

Before addressing the role of dissipative directionality in the stabilization of a current,
it is useful to revisit how a gauge field can naturally induce a flux-based current
in a ring geometry. The textbook case of a ground-state current in a closed system,
induced by a $U(1)$ gauge field, is revised in App.~\ref{app:ground_state_current}.
Here we discuss the analogous case of a flux-induced current, in the additional presence
of an (unstructured) environment.
One might indeed wonder whether it is possible to achieve a similar flux-based current in an open system, as well.

{Let us define the flux $\phi$ that pierces the ring by taking a complex tunneling strength
  \begin{equation}
    J = J_0 e^{i\phi/L}
  \end{equation}
  in the Hamiltonian of Eq.~\eqref{eq:hamiltonian}, where $L$ is the size of the system and $J_0$ is a positive constant.
We focus} on the simplest form of local dissipation, that is through
the addition of a local Lindblad jump term $L_j \propto \sigma^{-}_j$ on each site of the ring.
Since the net effect of incoherent spin flips along the $z$-axis is to polarize the steady state,
the only presence of a spin-hopping term in the Hamiltonian [first term in the r.h.s. of Eq.~\eqref{eq:hamiltonian}]
is not sufficient to counterbalance this effect, and thus the steady state would not carry any current
In order to create spin excitations, we thus add a non-zero $\lambda$ pairing term
[second term in Eq.~\eqref{eq:hamiltonian}].
The chemical potential $\mu$ fixes an energy scale to such excitations.
An analysis similar to the one for the ground-state current
leads to the emergence of a flux-dependent periodic current and a finite bipartite negativity,
as displayed in the left panel of Fig.~\ref{fig:steady_state_local_dissipation_phi_kappa}.
We observe a periodic current and a slightly changing negativity, but, as
for the case discussed in App.~\ref{app:ground_state_current}, 
there is no clear/direct relation between the two quantities.
As is the case for the ground state, this steady state current vanishes in the limit $L \to \infty$.
Note that, due to the presence of the pair creation/annihilation term $(\sigma_j^+\sigma_{j+1}^+ + {\rm h.c.})$
that breaks the particle conservation, the periodicity of the current is the double with respect to the closed case.

We conclude this section by studying the behavior of the steady state current
and of the bipartite negativity as function of the dissipation strength $\kappa$
and the gauge flux $\phi$ [see, respectively, middle and right panel
of Fig.~\ref{fig:steady_state_local_dissipation_phi_kappa}].
We observe the expected symmetry in the current around $\phi=\pi/2$, as well as a decrease
of the current with increasing $\kappa$. The negativity displays small values for small $\kappa$ (when the current is big),
then increases up to a maximum around $\kappa \approx 5$, and finally decreases again for very strong dissipation.
This behavior is mildly influenced by the value of $\phi$, but washed out for bigger values of $\kappa$,
since in the limit of strong dissipation, no excitations are left and the state is almost completely aligned along $z$.
It should be noted that the values of negativity are definitely smaller than the one observed for the ground-state current.
The resulting picture is that the (weak) entanglement present here is not related to the flowing current.

\section{Currents due to reservoir engineering}
\label{sec:CurrRes}

We now turn our attention to the particle current genuinely generated by the presence of the non-local
dissipation scheme.
We first address a minimal model that is able to sustain this kind of dissipatively induced current
(Sec.~\ref{sec:mean_field} and~\ref{sec:numerics}), and only later in Sec.~\ref{sec:perturbations}
we will consider the effect of adding other Hamiltonian, as well as dissipative terms.

To this purpose, the system's Hamiltonian can be assumed to be a simple chemical potential term,
such that $\mu$ fixes the energy scale.
As for the coupling of the bath, we only take the non-local dissipation $f_j(\{\sigma\})$, dropping the local terms.
To further reduce the amount of parameters, we set $\alpha=\gamma^*$ and $\delta=\beta^*$.
In this circumstance, the magnetization along $z$ reads:
\begin{equation}
  \label{eq:magnetization_non_local_dissipation}
  \dfrac{\mathrm{d}}{\mathrm{d}t} \sigma^{z}_{j} = - \big(\sigma^{z}_{j}+1\big) \Gamma^{-} - \big(\sigma^{z}_{j}-1\big) \Gamma^{+}
  - \big( {\cal I}^{\eta}_{j} + {\cal I}^{\eta}_{j+1} \big) ,
\end{equation}
where we now have $\eta =\alpha^{2}-\delta^{2}$ as the coupling constant to the current,
$\Gamma^{-} = 2|\delta|^2$ and $\Gamma^{+} = 2|\alpha|^2$.
This leads to a current ${\cal I}^{\eta}_{j} = \big( -\eta \, \sigma^{+}_{j}\sigma^{-}_{j-1} + \textrm{h.c.} \big)$.

Each of the first two terms in the r.h.s.~would drive the system towards either
$\langle\sigma^{z}_{j}\rangle=1$ or $\langle\sigma^{z}_{j}\rangle=-1$, respectively.
If there is a frustration between the two of them, the other terms in Eq.~\eqref{eq:magnetization_non_local_dissipation}
will mend this and a current will ensue.
This also means that for $\delta=0$ (or $\alpha=0$) such a frustration does not occur
and the system is driven into a steady state which is completely aligned (or anti-aligned) along the $z$ axis,
without any current flow.
Furthermore, if $\alpha=\delta$ the two terms cancel out, without the need for a current to arise.
Indeed, in such case the steady state is a completely mixed state,
as the Lindblad operators create as many excitations as they annihilate.
In all the other cases, the imbalance between creation and annihilation of excitations naturally generates a current.
Lastly, we observe that, as currents emerge from the structured reservoir and its induced imbalance in the system,
the current studied here is not a conserved one, as is the case in the ground-state scenario.

\subsection{Mean-field treatment}
\label{sec:mean_field}

To get more insight into the physics of the dissipative model, we have developed a CMF treatment
in the spirit of Refs.~\cite{Tomadin_2010, Jin_2016}, where the system's density matrix is supposed
to be written in a cluster-factorized form:
\begin{equation}
  \rho_{\rm CMF} = \bigotimes_{\cal C} \rho_{\cal C}.
  \label{eq:CMFansatz}
\end{equation}
Here we consider the simplest case which admits the possibility to have directionality,
that is an ansatz where each cluster ${\cal C}$ is made of two neighboring spins, and then study the current,
the on-site magnetization, the negativity and the purity of the state $\rho_{\cal C}$.
In practice, we restrict to a subsystem ${\cal C}$ of the global system containing only two sites,
and treat the interaction with the outer part of system by means of a mean-field variables.
We then solve self-consistently the dynamic equation for $\rho_{\cal C}$, with respect to these mean-field parameters.
Notice that, while typical applications of CMF methods replace parts of the Hamiltonian by a mean-field variable,
here we need to perform an analogous decoupling on the dissipative part acting outside the two-site cluster.

Since the model features a global spin flip symmetry, the two-site density matrix has a so-called ``X-structure'',
which considerably simplifies the calculations (see, e.g., Ref.~\cite{Amico_2004}).
Here we report the relevant results, while further details
are postponed to App.~\ref{app:mean field}.
We find that the only non-zero current is the circulating contribution:
\begin{equation}
  \langle {\cal I}^\eta \rangle = \!\dfrac{4\alpha^2\delta^2(\delta^2-\alpha^2)^3}
  {(\alpha^2-\delta^2)^2 (\alpha^4+\delta^4+3\alpha^2\delta^2)+\mu^2(\alpha^2+\delta^2)^2}.
  \label{eq:Curr_CMF}
\end{equation}

\begin{figure}[!t]
  \includegraphics[width=0.49\columnwidth]{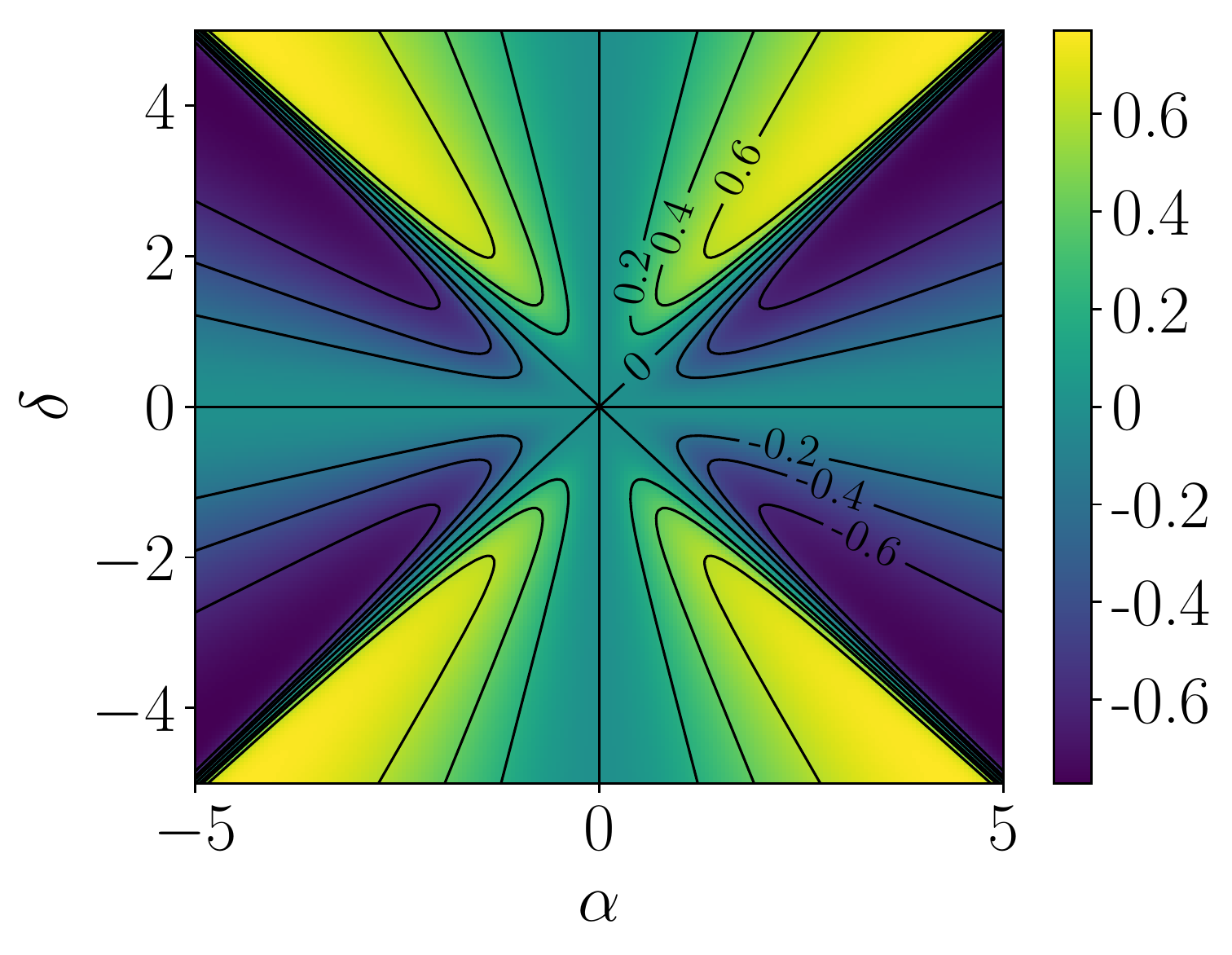}
  \includegraphics[width=0.49\columnwidth]{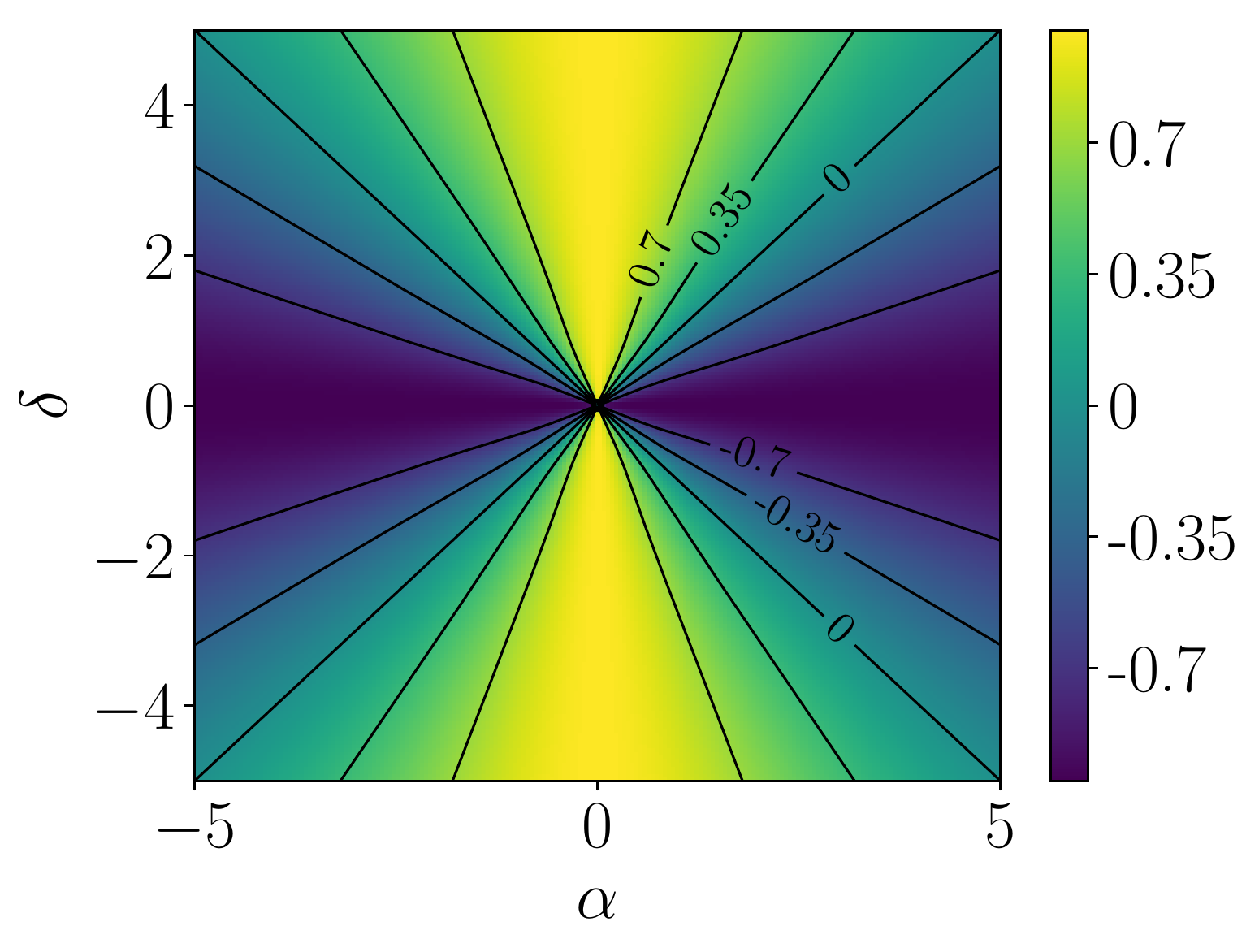}
  \includegraphics[width=0.49\columnwidth]{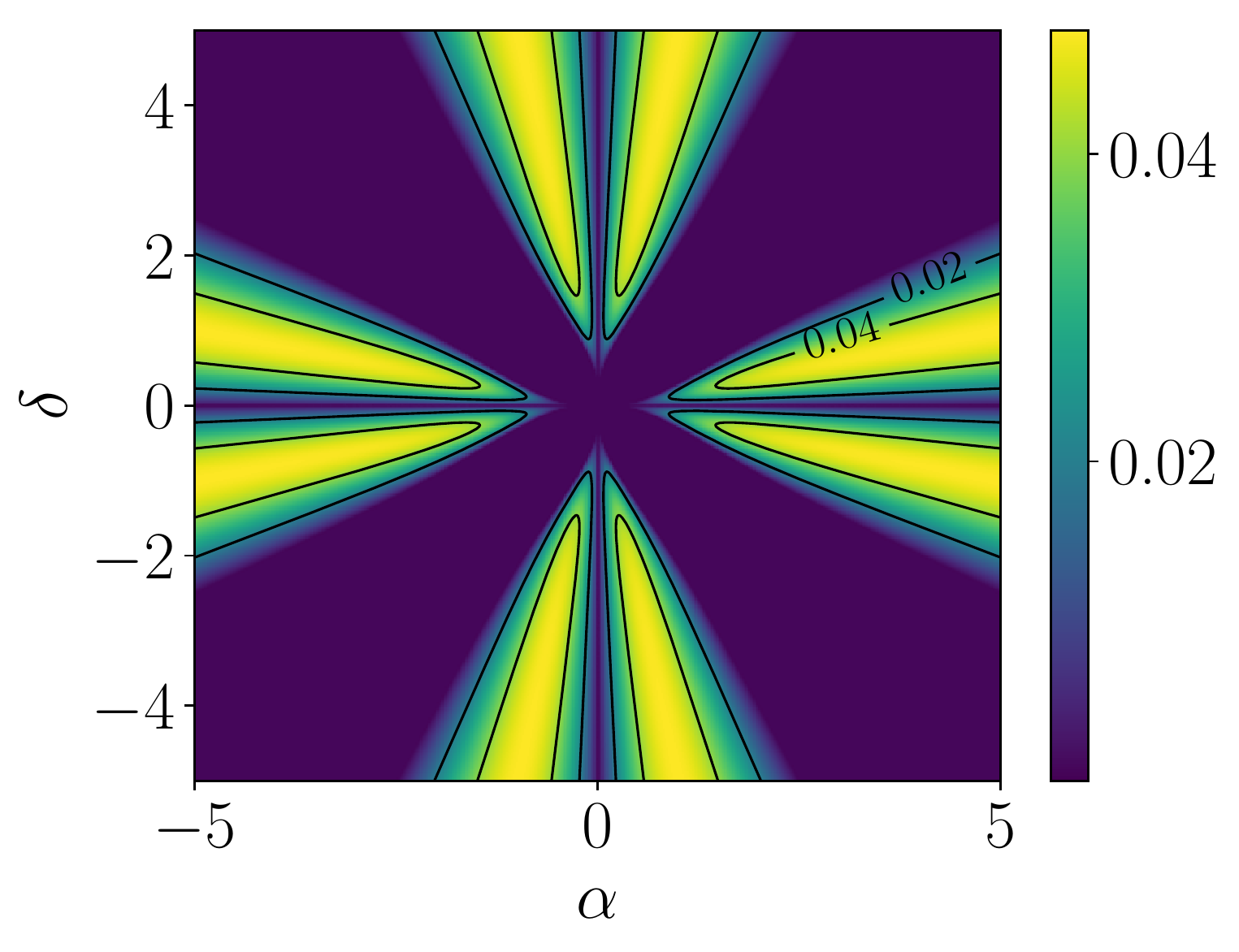}
  \includegraphics[width=0.49\columnwidth]{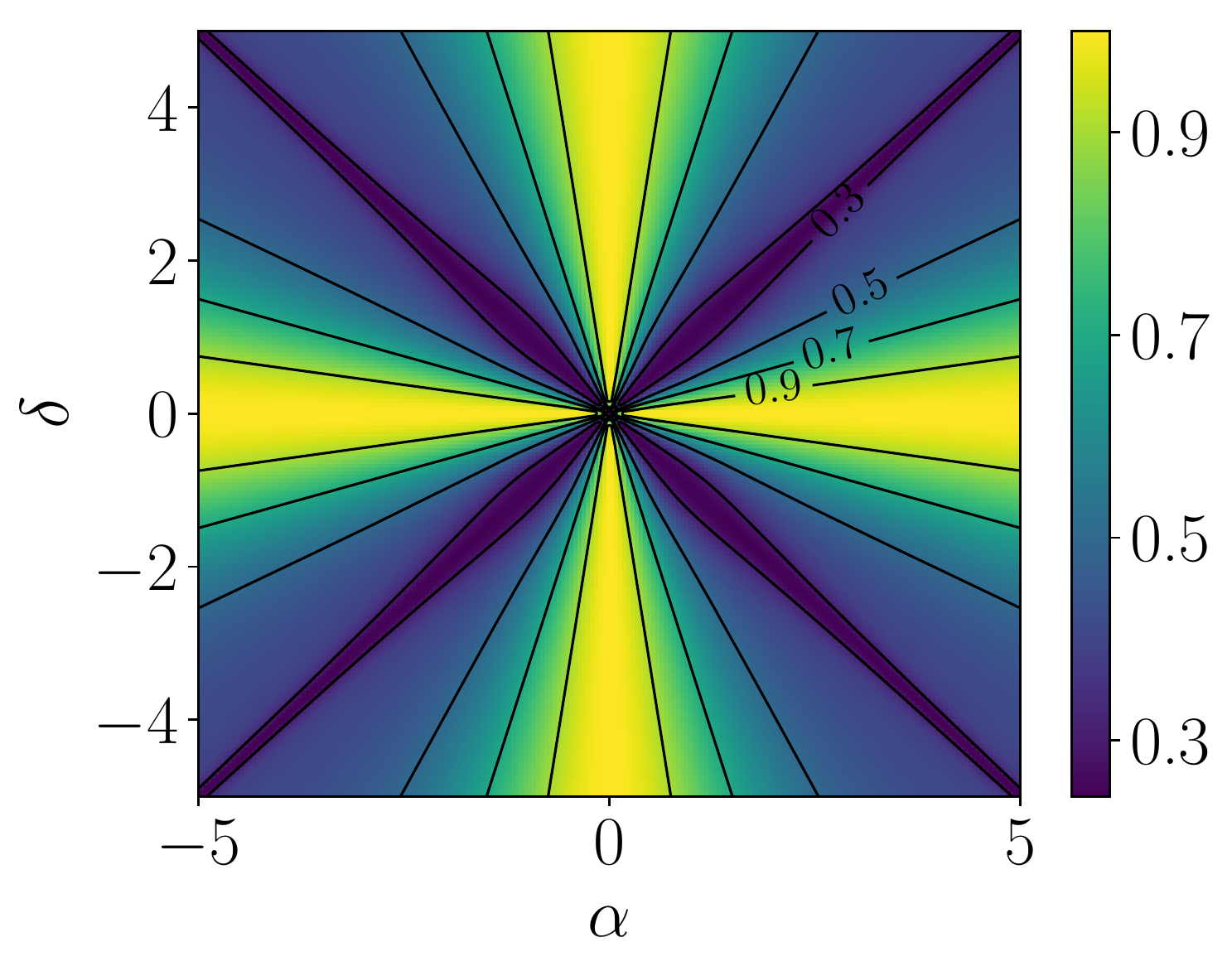}
  \caption{Normalized, dissipatively induced current $\langle {\cal I}^\eta \rangle/|\alpha^2-\delta^2|$ (top left),
    $z$-axis magnetization $\langle \sigma^z \rangle$ (top right),
    negativity $N(\rho_{\rm SS})$ (bottom left), and purity $P(\rho_{\rm SS})$ (bottom right)
    as a function of the two parameters $\alpha$ and $\delta$ of the engineered dissipation.
    Data have been obtained through a CMF approach applied to our minimal scheme of non-local dissipation.}
  \label{fig:meanfield}
\end{figure}

Figure~\ref{fig:meanfield} provides some results obtained from the ansatz in Eq.~\eqref{eq:CMFansatz}.
Since the dissipator scales with the dissipation strength $\alpha^2-\delta^2$,
in the top left panel we have plotted the normalized quantity $\langle {\cal I}^\eta \rangle /|\alpha^2-\delta^2|$.
We observe that the symmetries of the system are being respected in all of the plots.
It is interesting to notice that, when $\alpha$ and $\delta$ are almost equal,
the current takes very big values, but at the same time it is very sensitive to changes in the parameters:
a small change in them makes the current sharply change from positive ($\alpha>\delta$)
to negative ($\alpha<\delta$) values.
Further, in all quantities we observe the symmetries with respect to $\alpha$ and $\delta$.
The magnetization $\langle\sigma^{z}\rangle$ (top right) follows the trend expected from the current:
if the system holds no current, it is maximal ($\langle \sigma^z \rangle = \pm1$);
conversely, if the current is maximal, the magnetization is zero.
The negativity (bottom left) and the purity (bottom right) show slightly correlated behavior with the current,
in particular large-current states have zero negativity and are highly, but not completely, mixed.
States carrying no current are pure and without any negativity, as they are fully (anti-)aligned along $z$.

\subsection{Numerical results}
\label{sec:numerics}

\begin{figure*}[!t]
  \includegraphics[width=0.9\columnwidth]{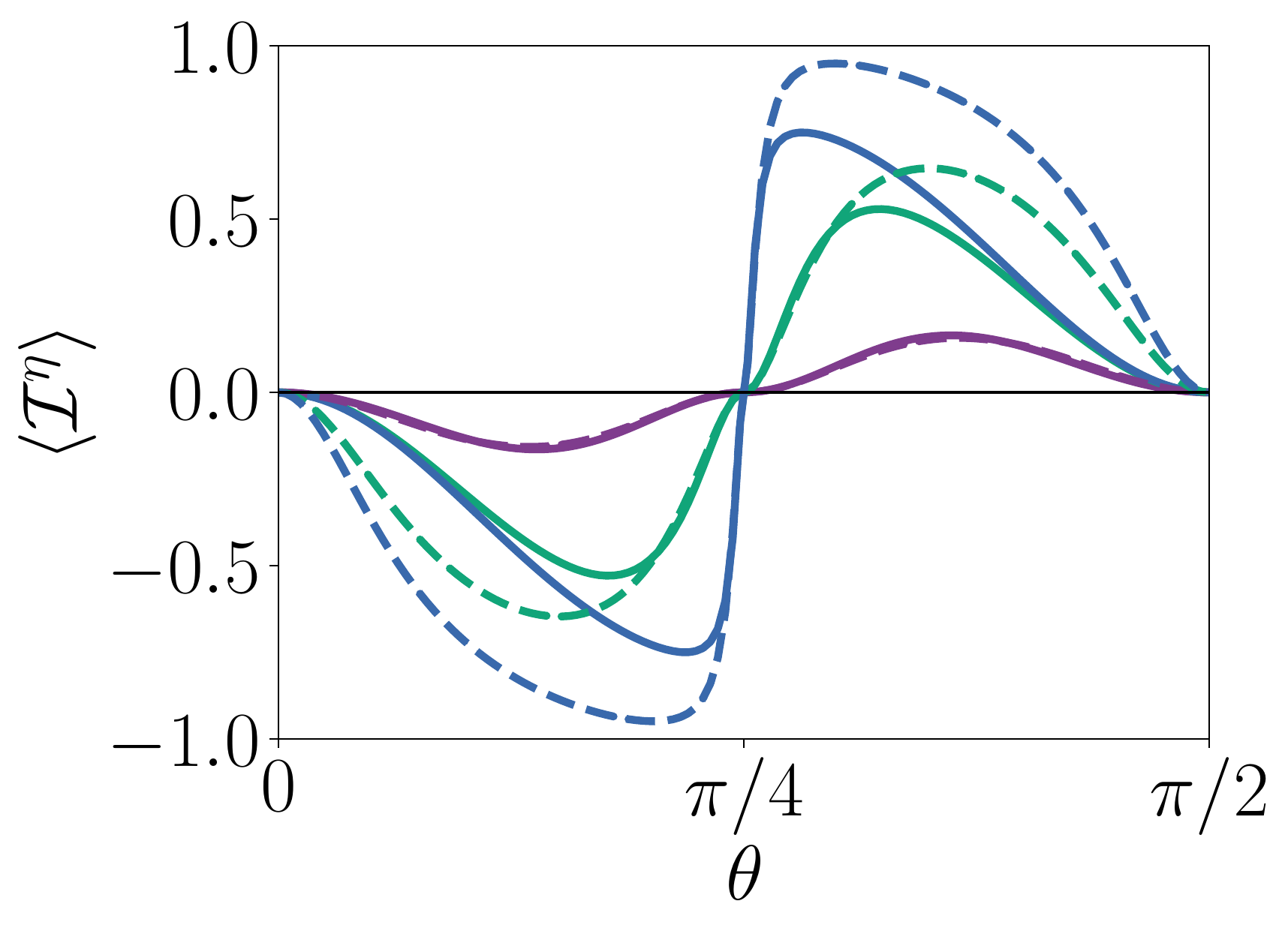}
  \hspace{0.75cm}
  \includegraphics[width=0.9\columnwidth]{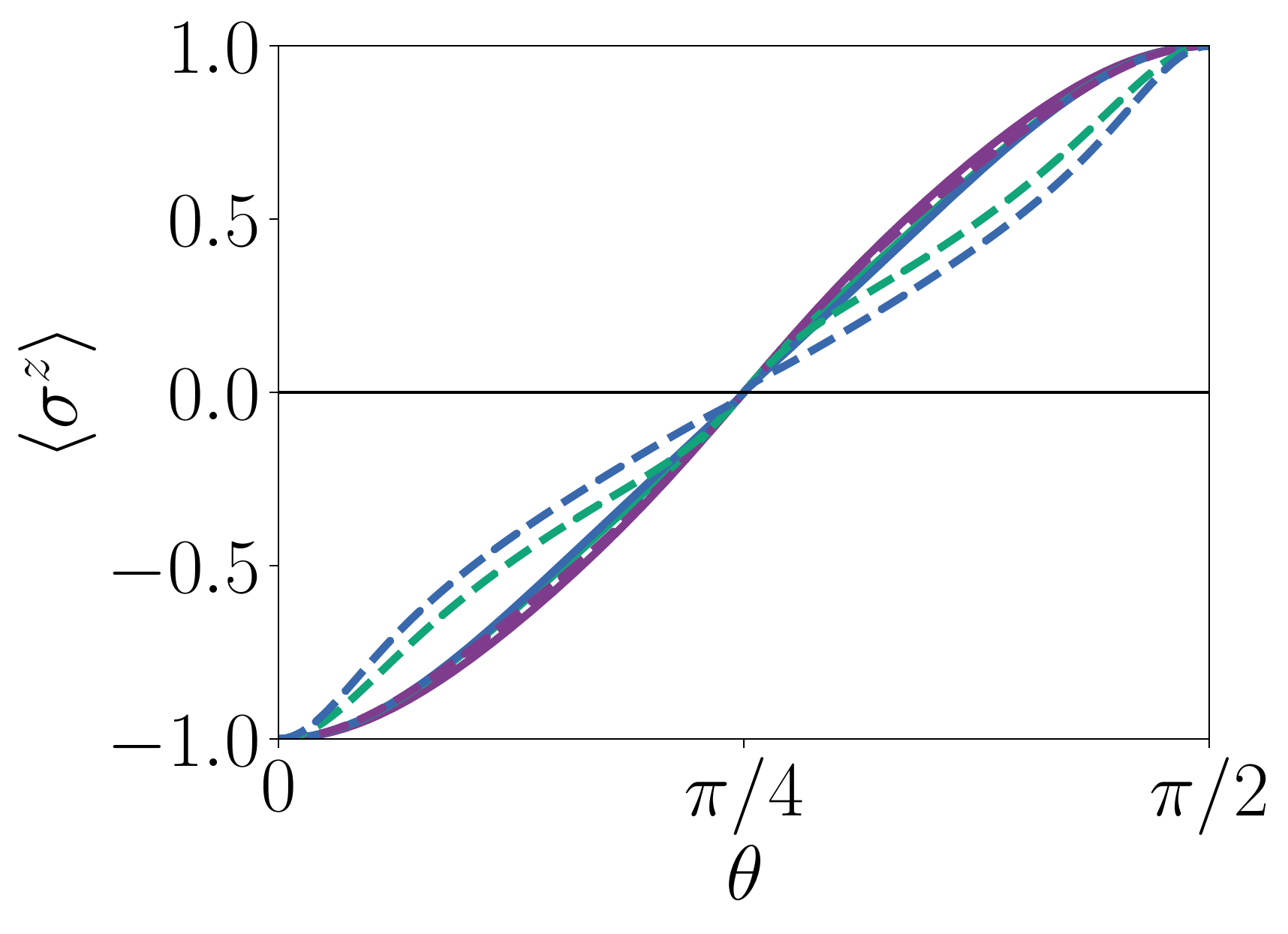}

  \vspace{0.25cm}
  \includegraphics[width=0.9\columnwidth]{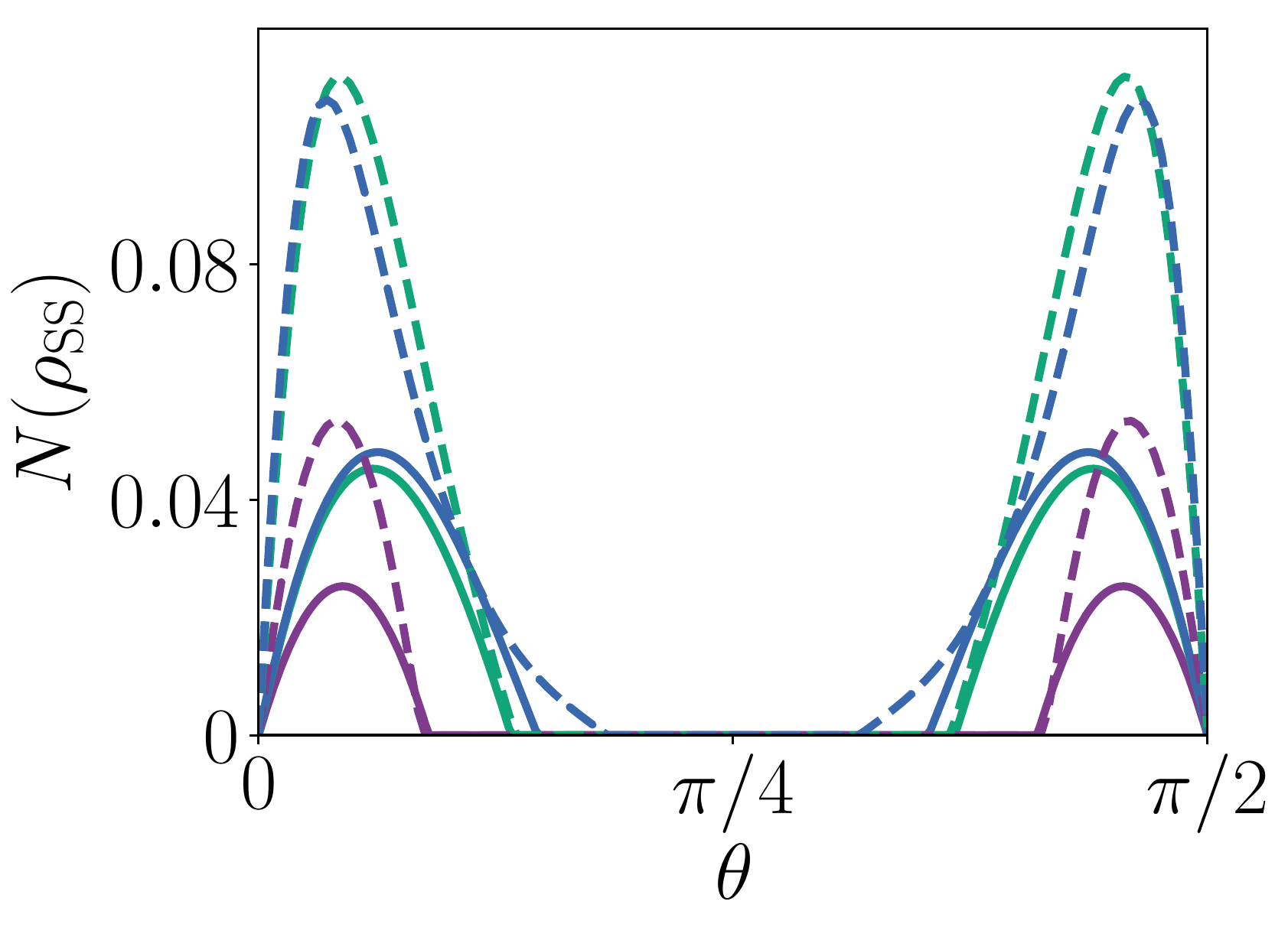}
  \hspace{0.75cm}
  \includegraphics[width=0.9\columnwidth]{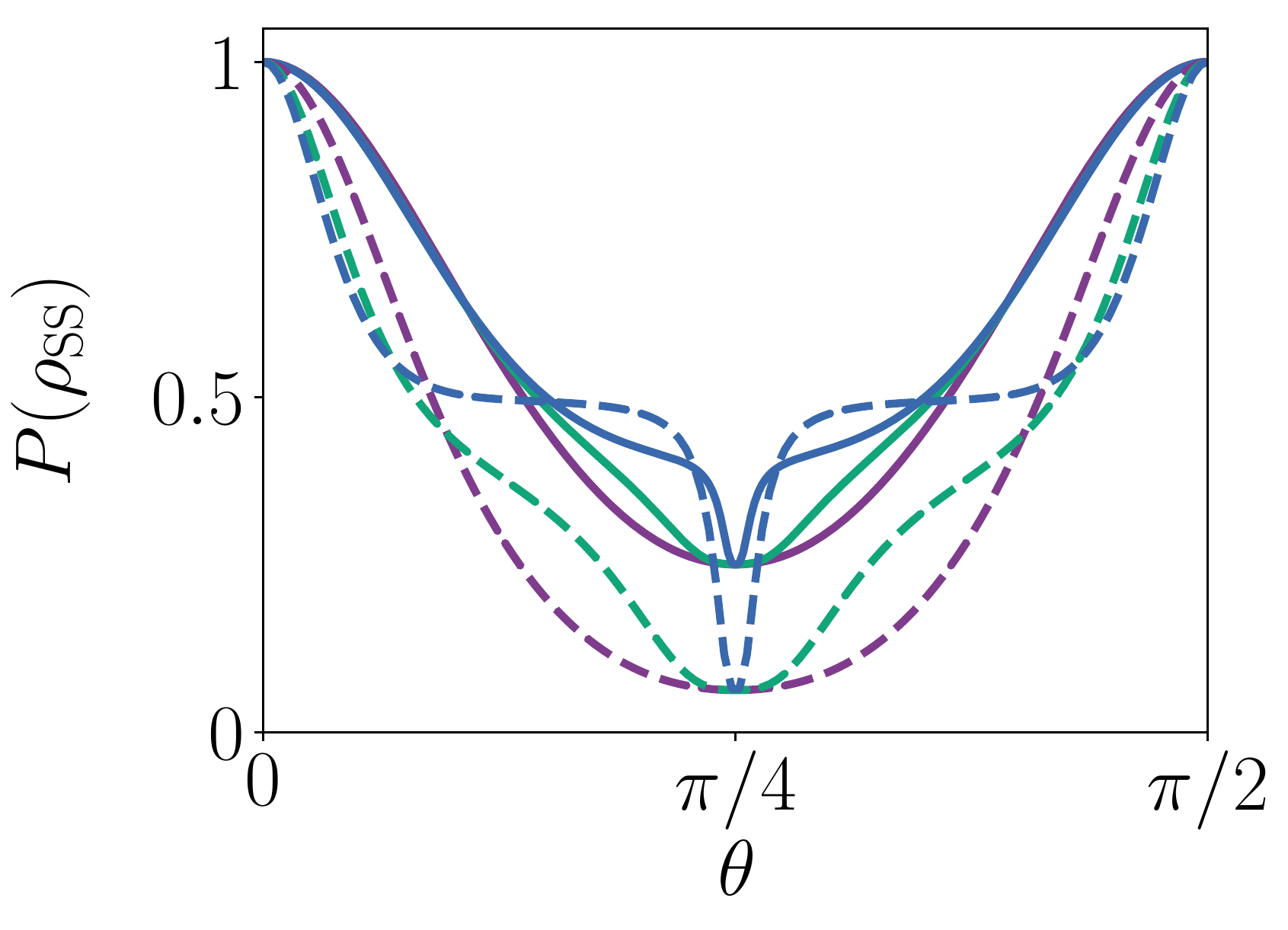}
  \caption{Comparison between CMF and ED results for $L=4$. 
    The dissipation parameters are chosen as $(\alpha,\delta)=r\,(\cos \theta,\sin \theta)$,
    The various panels show the behavior of several quantities as functions of $\theta$:
    dissipatively induced current (top left), magnetization along $z$ (top right),
    negativity (bottom left) and purity (bottom right).
    The various curves correspond to: $r=1$ {(dark magenta)}, $r=2$ {(light green)} and $r=5$ (blue),
    while solid lines are the results by ED and the dashed lines are from the CMF calculations.}
  \label{fig:comp}
\end{figure*}

We complement the mean-field analysis by a numerical investigation of the system, using
exact diagonalization (ED), for systems with up to $L=8$ sites, and MPOs for larger system sizes (up to $L=30$).
To assess the quality of the mean-field result, we have chosen to parametrize $\alpha$ and $\delta$
as $(\alpha,\delta)=r\,(\cos \theta,\sin \theta)$.
Some results for $L=4$ are displayed in Fig.~\ref{fig:comp}. We observe that, for values of $r\approx \mu$,
the CMF results for the current (upper left panel) are in remarkable qualitative agreement with the ED.
As $r$ is increased, the quantitative agreement between the two prediction generally diminishes.
Such observation qualitatively holds also for all the other quantities that we monitored
(magnetization, negativity, and purity --- see the other three panels).
This leads to the conclusion that CMF is only able to catch all the details in the ``weak dissipative'' regime ($r/\mu \lesssim 1$),
while in the ``strong dissipation regime'' ($r/ \mu \gg 1$) the agreement becomes more qualitative
and only some features are captured by CMF.

\begin{figure}[!t]
  \includegraphics[width=0.8\columnwidth]{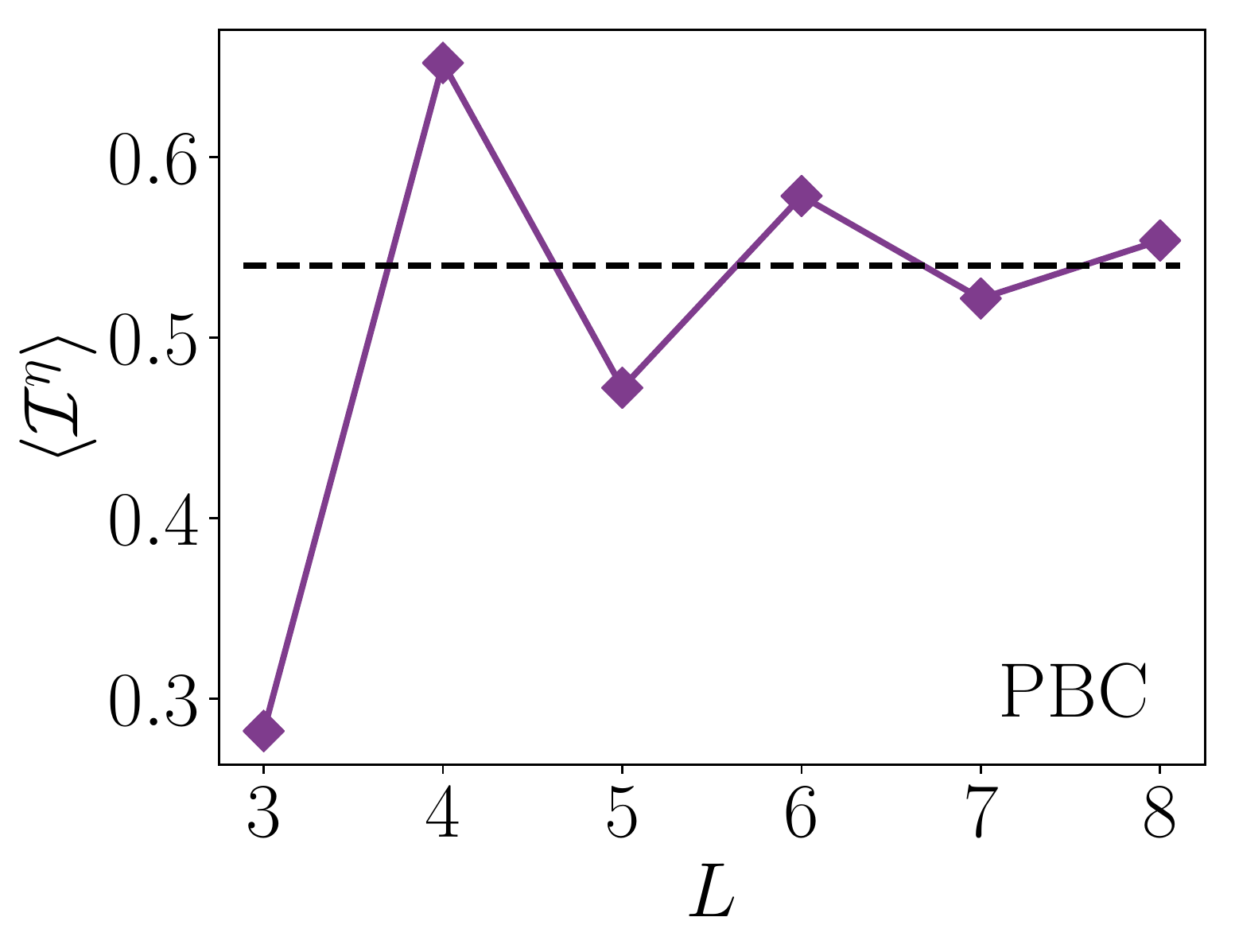}
  \includegraphics[width=0.8\columnwidth]{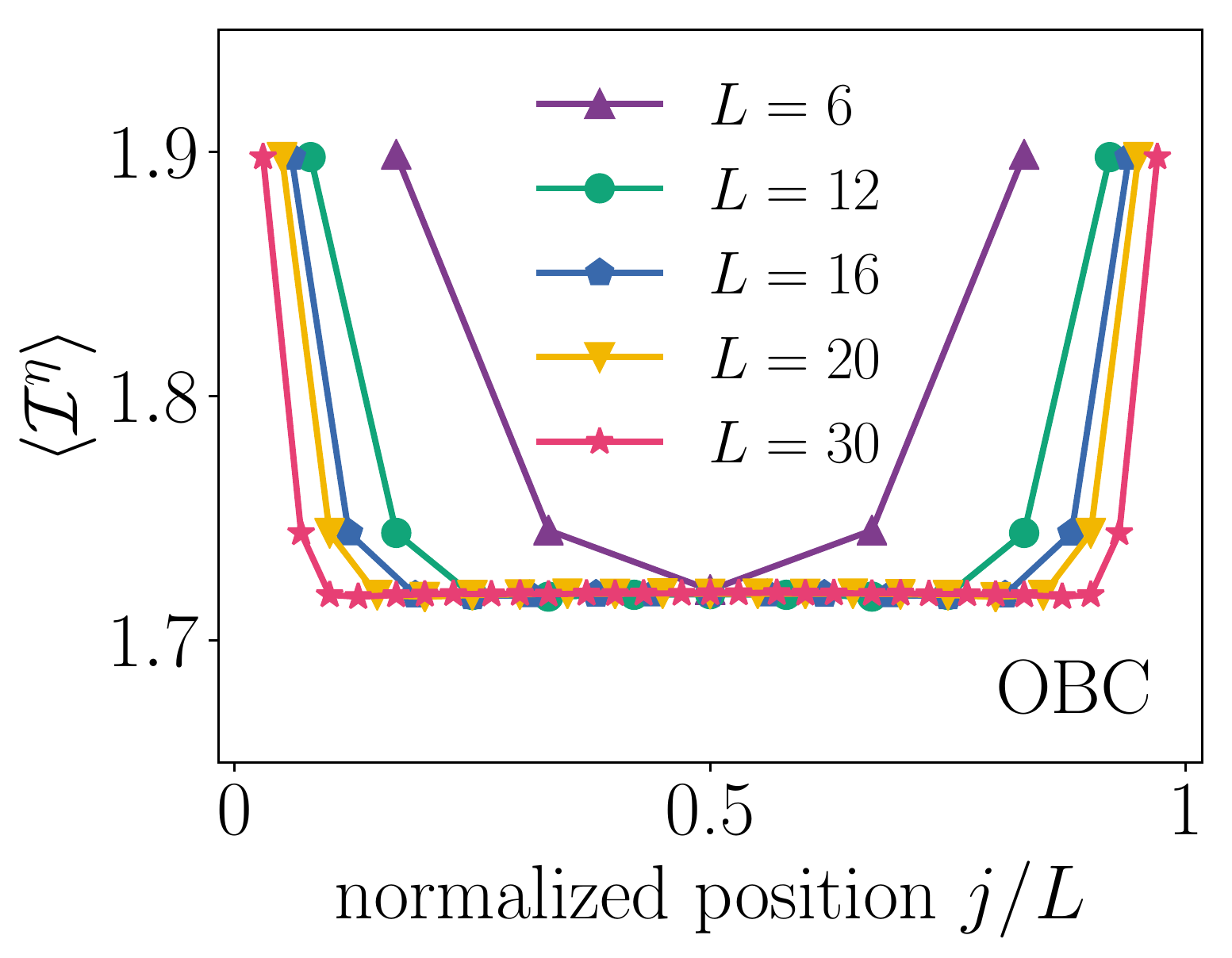}
  \caption{Numerical results for the dissipative current $\langle {\cal I}^{\eta} \rangle (L)$ as a function
    of the size $L$.
    Upper panel: scaling of the dissipative current with $L$ for a periodic ring.
    The alternating character supports the trend of insetting convergence (in $L$) to a finite,
    non-zero value of the steady state current. Here we fixed $\alpha = 2$ and $\delta = 1.5$. 
    Lower panel: dissipatively induced current for an open chain as a function of normalized position $j/L$.
    The various data sets are for different system lengths $L$; the approach to a finite current in the bulk is evident.
    Here we fixed $\alpha = 2$ and $\delta = 1$.}
  \label{fig:numerics}
\end{figure}

By scaling the dimension of the system up to $L=8$, it is possible to observe {an indication of} convergence
of $\langle {\cal I}^\eta \rangle (L)$
toward a finite value at large values of $L$ (left panel of Fig.~\ref{fig:numerics}),
thus proving that reservoir-based currents are sensibly different from currents induced by gauge fields
(the latter representing a boundary effect, which valishes in the thermodynamic limit).
Specifically, our numerics evidences the emergence of an alternating character in the current
$\langle {\cal I}^\eta \rangle (L)$ between odd and even sites, as well as a (rapidly) insetting convergence.

The system size can be further increased by adopting a MPO approach~\cite{MPS_dissipative}, and focusing on
systems with open boundary conditions. Our results in Fig.~\ref{fig:numerics} show that
the bulk clearly supports a finite value of the current $\langle {\cal I}^\eta \rangle (L)$,
as in the case of a closed ring, thus spotlighting the macroscopic persistence of such current.
Our expectation that the bulk of a chain of sufficient size shows a current similar to the periodic-case result,
is indeed verified by studying its the convergence with $L$.
We note the presence of an approximately constant bulk current that qualitatively agrees with the current obtained in a ring.
{These observations, together with the success of the mean-field ansatz in describing this phenomenon,
  suggests the stabilization of a persistent current to a finite value for $L\rightarrow\infty$, irrespective
  of the choice of boundary conditions.}

\subsection{Perturbations}
\label{sec:perturbations}

We end up our study with a discussion of the effects induced on the dissipative current
by the presence of possible perturbations/extensions to the minimal model presented before.
Namely, we consider perturbations in the form of:
{\it i)} localized dissipation; {\it ii)} on-site and nearest-neighbor Hamiltonian terms;
{\it iii)} replacing the nearest-neighbor dissipation by next-to-nearest-neighbor dissipation,
and study the effects of this on the current.

Adding a local dissipation $\kappa\, \sigma_j^-$ on each site amounts to replacing $\Gamma^- = 2\, |\delta|^2$ by $\Gamma^- + \kappa > 0$.
As a consequence, this breaks the symmetry between $\alpha$ and $\delta$ and allows for a non-zero current even for $\delta=0$.
Due the simply rescaling of $\Gamma^-$, no further effects arise; the current is stable with respect
to this perturbation as it keeps its main characteristics.

\begin{figure}[!t]
  \includegraphics[width=\columnwidth]{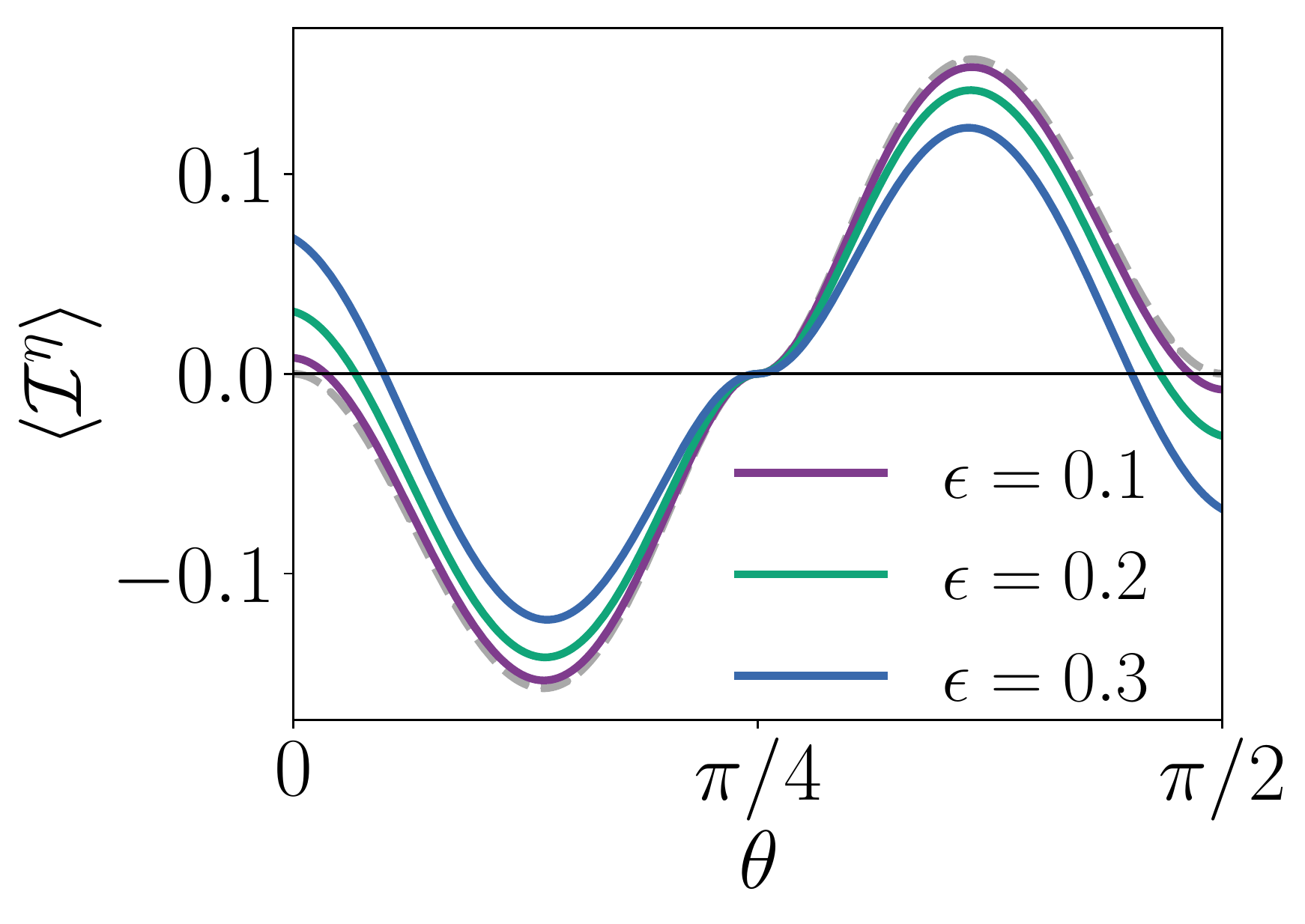}
  \includegraphics[width=\columnwidth]{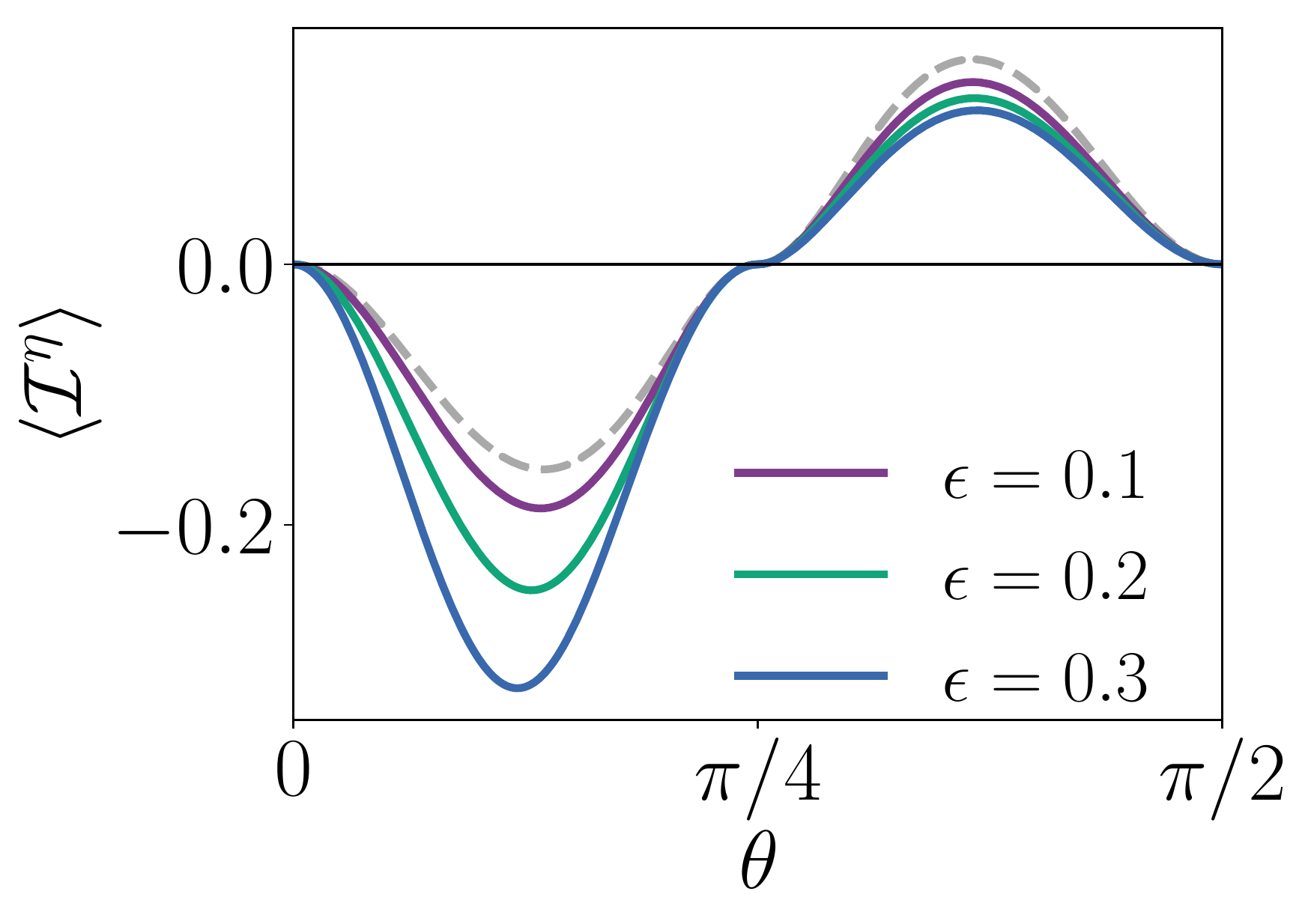}
  \caption{Dissipatively induced current as a function of the parametrizing angle $\theta$, at fixed $r=1$,
    in the presence of perturbations of our minimal model.
    The grey {dashed} line shows the reference value for an unperturbed system.
    Upper panel: local perturbation by a longitudinal field $\sigma_j^x$.
    Lower panel: nearest-neighbor perturbation by an interaction term $\sigma_j^z\sigma_{j+1}^z$.
    The various curves are for different values of the {perturbation strength $\epsilon$:
      $\epsilon=0.1$ (dark magenta), $\epsilon=0.2$ (light green) and $\epsilon=0.3$ (blue).}}
  \label{fig:Hamiltonian_perturbation}
\end{figure}

Now we consider the possibility to include more Hamiltonian terms in the model.
We start by adding a local term $\epsilon\, \sigma_j^x$ and study the behavior of the current
via the parameterization used before, $(\alpha,\delta)=r\,(\cos \theta,\sin \theta)$.
In the left panel of Fig.~\ref{fig:Hamiltonian_perturbation}, we show the current as a function of $\theta$,
for $r=1$ and various values of the perturbation strength $\epsilon$.
We observe that the symmetry around $\theta=\pi/4$ is preserved, but the maximum value of the current
is damped with increasing $\epsilon$, while at the same time there is a non-zero current for $\theta=0,\pi/2$ that increases with $\epsilon$.
These features are readily explained: the additional term polarizes the spins partially along the $x$ direction,
making it less susceptible for the dissipative mechanism, but at the same time this breaks the symmetry
between $\alpha$ and $\delta$, such that even if one of them is zero, a current can however flow.
Apart from this, the general behavior remains the same: the persistent current is a main feature also for this extended model.

An equally relevant and straightforward extension is to add nearest neighbor interaction.
Here we choose a term $\epsilon\, \sigma_j^z\sigma_{j+1}^z$.
The right panel of Fig.~\ref{fig:Hamiltonian_perturbation} again displays the current as function of the angle $\theta$,
for various values of $\epsilon$. We observe that also for non-zero $\epsilon$ the current is zero for $\theta = 0, \pi/4, \pi/2$.
Additionally, the symmetry around $\theta = \pi /4$ is broken: the current is damped for $\theta \in [\pi/4,\pi/2]$,
while it is increased for $\theta \in [0,\pi/4]$.
This can be understood as follows: depending on the sign of $\langle\sigma_{j+1}^z\rangle$, the added term
drives the system toward $|\hspace{-.3em}\uparrow~\!\!\!\rangle$ or $|\hspace{-.3em}\downarrow~\!\!\!\rangle$,
thereby effectively enhancing one of the two dissipative parameters $\Gamma^\pm$, yielding a shift in the current.

Lastly, we discuss the consequences of a dissipation being engineered such that it affects next-to-nearest neighbors
on a current flowing between nearest neighbors.
If we consider an even amount of spins, we directly see that this dissipation splits the system in two separate systems:
one for the odd sites and one for the even sites. As a consequence no current flows between nearest neighbors.
A different situation can be found for a chain with an odd number of spins, since here the complete system is connected.
Therefore, a current between nearest neighbors is established. However, nearest neighbors are only connected
through several dissipative terms going around the whole chain before reaching the neighboring site.
As a consequence, it is much smaller in magnitude than the nearest-neighbor induced current,
and in the limit $L\to\infty$ it vanishes. This eventually merges the odd and the even result, as expected. 

Concluding, we have seen that the current studied in our minimal model is not a peculiarity of the model,
but is robust when perturbed by additional Hamiltonian terms or when additional, next-to-nearest neighbor dissipation is present.

\section{Conclusions}
\label{sec:conclusion}

In summary, we have introduced a dissipative setting to engineer macroscopic, persistent steady-state currents. 
We have studied them by means of a cluster mean-field treatment, resulting in a simple expression for the current,
and supported our predictions with exact numerical calculations.
A qualitative agreement emerges for all the parameter regimes that we considered;
this becomes even excellent in the regime of weak dissipation.
Furthermore, we have considered the behavior of the current for larger systems,
and showed that it is non-vanishing as the system size increases.
Additionally, we have investigated how the current is influenced by local imperfections,
demonstrating that it is robust with respect to them.

When comparing the different kinds of current, i.e.,  those induced by gauge fields and those due to non-local engineered dissipation,
we observed that the dissipatively induced current is related to the negativity, while such a phenomenon
is not observed in the other case. This could be a signal that the mechanism behind it has a genuine quantum nature.
However, as the cluster mean-field approach leads to qualitatively correct results, this is not expected to play the main role.
The striking difference is that the dissipative current is not vanishing in the macroscopic limit $L \to \infty$, while, per construction,
the gauge field can be seen as a boundary term which leads to a vanishing current for large sizes.

The work presented here constitutes the first and simplest situation where it is possible to establish
a mechanism of reservoir-induced currents in 1D ring-shaped lattices of QED cavities.
It is worth mentioning that recent advances in quantum simulation enabled to realize lattices of more complex topology,
where non trivial flux-induced currents can be generated.
For example, it has been possible to experimentally realize the so-called Hofstadter butterfly~\cite{Bloch_2013, Ketterle_2013},
by engineering cold atomic systems which mimic the behavior of quantum matter in two-dimensional crystalline structures,
in the presence of strong magnetic fluxes induced by laser-assisted tunneling mechanisms.
Other experiments exploited the internal atomic degrees of freedom as a synthetic dimension,
in order to realize effective ladder geometries with physical 1D systems~\cite{Fallani_2015, Spielman_2015}.
Lastly, there has been a considerable effort in the study~\cite{Peano_2015}
and realization~\cite{Barik_2018} of hybrid light-matter devices,
that have the potential to combine positive properties of various platforms for quantum technologies.
These hybrid systems also provide the chance to study higher dimensional systems,
where quantum Hall-like physics can be realized~\cite{Ozawa_2018}.
It is not difficult to imagine that, following the same path, it will be possible to design frustrated systems
or higher dimensional topological states using reservoir engineering.

\acknowledgments

We acknowledge Fernando Iemini for useful discussions. This work is partly supported by EU-IPQUIC.

\appendix

\section{Dissipation-induced directionality}
\label{app:BH model}

Here we provide some details on the calculation of the particle current, focusing on the bosonic model
of Eqs.~\eqref{eq:HamModel}-\eqref{eq:Diss_def}. We focus on the equation of motion,
as well as on a more detailed discussion on directionality and non-reciprocity.
First of all, the Lindblad equation for the density matrix, Eq.~\eqref{eq:MasterModel},
can be recast as an equation of motion for a generic observable $A$ as
\begin{equation}
  \dfrac{\mathrm{d}}{\mathrm{d}t} A = i [H,\rho]
  + \sum_{j} \kappa_{j} \, \mathcal{\hat D}[d_{j}] \, A + \sum_{j} \mathcal{\hat D}[f_{j}(\{\sigma\})] \, A, 
\end{equation}
where
\begin{equation}
  \mathcal{\hat D}[O] \, A = O^\dagger A O - \tfrac{1}{2} \{O^\dagger O, A\} .
\end{equation}
Making use of the standard bosonic commutation relations $[d_n,d^\dagger_m]=\delta_{n,m}$ and $[d_n,d_m]=0$,
the equation of motion~\eqref{eq:EoM_BH} for the $d_j$ operator naturally follows.

In typical situations, as is the case in the absence of artificial gauge fields, interactions in quantum mechanics
are reciprocal. For a lattice system, this simply means that interactions to the left and to the right
are equal and, as such, no distinction between them is possible. While it is known that this symmetry
can be easily broken by inserting the aforementioned magnetic or artificial gauge field,
this is not the only way.
As a matter of fact, Eq.~\eqref{eq:EoM_BH} clarifies that not only the unitary dynamics created
by the term $J d^\dagger_j d_{j+1} + J^* d^\dagger_{j+1}d_j$, with $J$ complex, can differentiate
between left ($j-1$) and right ($j+1$), but also the two dissipative terms with effective coupling
constants $\eta$ and $\xi$ entail such a sensitivity.

This fact leads to a wealth of possibilities. For example, an appropriate choice of parameters
leads to the cancellation of certain terms, making it possible to get a unidirectional
equation of motion~\cite{Metelmann_2015}. In our example, this means that $d_j$ is influenced
from the left ($d_{j-1}$ is part of the equation of motion), but not from the right ($d_{j+1}$ does not appear
in the equation of motion), as is the case in Eq.~\eqref{eq:EoM_BH_unidirectional}.

An alternative route explored in this manuscript is to study the possibility of purely dissipative phenomena
breaking reciprocality. In this case, the unitary dynamics is either not breaking reciprocality
or not present at all. Since the latter case already unveils a plethora of features,
in this work we limit our study to that one.

\section{Ground-state current}
\label{app:ground_state_current}

In a closed system, the simplest theoretical scheme on a lattice that captures the physics of
a circulating particle current is a non-interacting tight-binding model on a 1D ring,
in the presence of a $U(1)$ gauge field. The physical mechanism that induces a current
is that of the Aharonov-Bohm effect.
In its simplest form, this is described by the Hamiltonian~\eqref{eq:hamiltonian},
with $\lambda=0$, $\mu=0$, in the presence of a complex tunneling strength $J = J_0 e^{i \phi /L}$,
that is:
\begin{equation}
  H = J_0 \sum_{j=1}^L \big( e^{i \phi/L} \sigma^{+}_{j}\sigma^{-}_{j+1} + \textrm{h.c.} \big) .
  \label{eq:Ham0}
\end{equation}
Here $\phi \in [-\pi,\pi]$ denotes the effective magnetic flux.
From the continuity equation for the magnetization along $z$,
one finds the associated current
\begin{equation}
  \label{eq:ground_state_current}
   {\cal I}^{J}_{j} = 2iJ_0 \big( e^{i\phi/L}\sigma^{-}_{j}\sigma^{+}_{j+1} - \textrm{h.c.} \big) ,
\end{equation}
which is in accordance with the Eq.~\eqref{eq:MagCur} specialized to this case.

\begin{figure}[!t]
  \includegraphics[width=0.8\columnwidth]{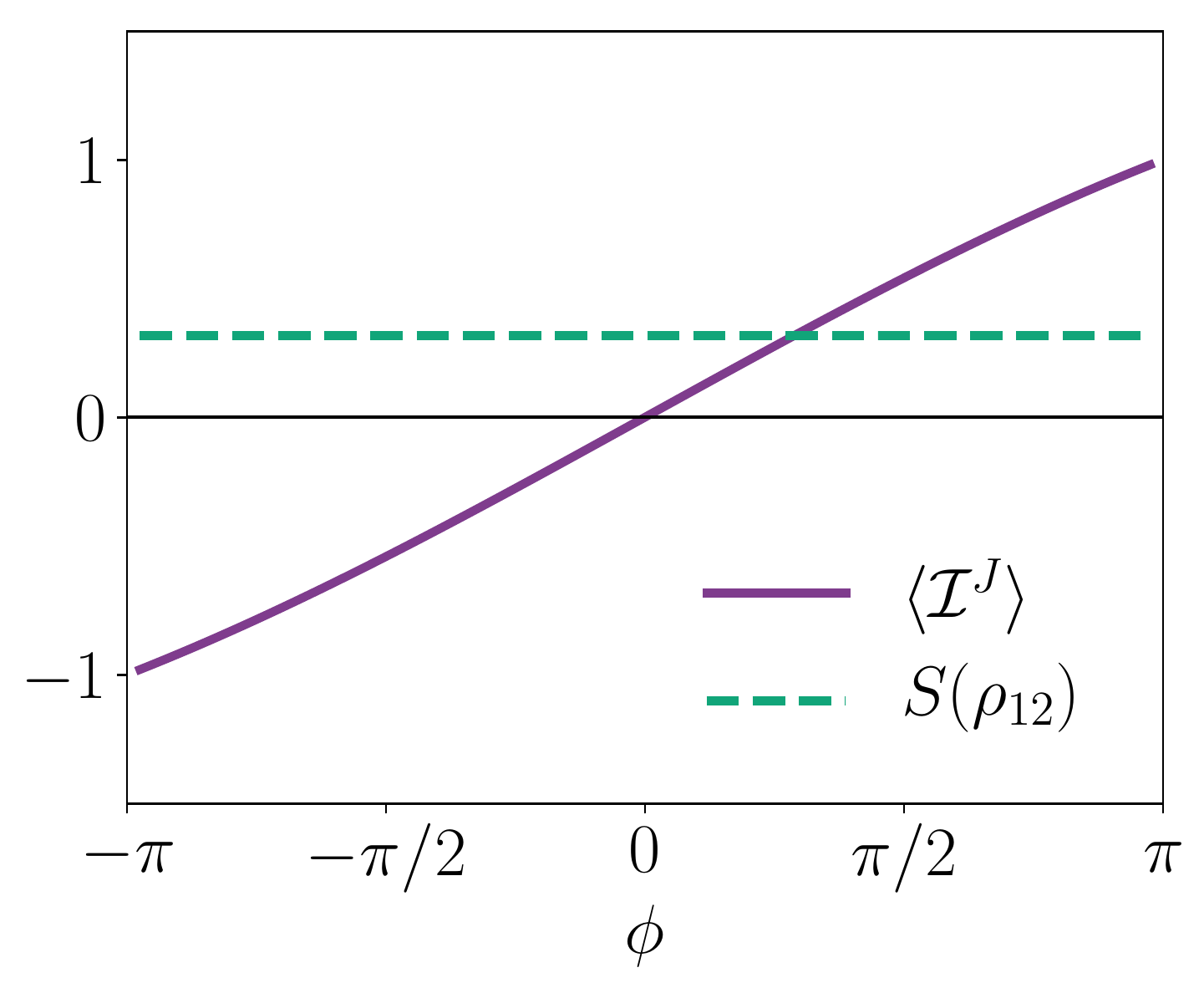}
  \caption{Flux-induced particle current in the ground state of a tight binding model of size $L=4$,
    described by Eq.~\eqref{eq:Ham0}.
    We plot the local current $\langle {\cal I}^J \rangle = \langle \psi_G| {\cal I}^J |\psi_G\rangle$,
    where ${\cal I}^J$ is given in Eq.~\eqref{eq:ground_state_current} with $J_0=1$ {(dark magenta, continuous line)},
    and the entanglement entropy $S(\rho_{12})$ (light green, dashed line) for the two-site reduced density matrix
    $\rho_{12} = {\rm Tr}_{3,4} \big[ |\psi_G\rangle \langle \psi_G| \big]$,
    as a function of the flux $\phi$.
    Here $|\psi_G\rangle$ denotes the ground state of the Hamiltonian.
    Notice that, due to translational invariance, the measured current is independent of the position $j$.}
  \label{fig:ground_state_current}
\end{figure}

The unitary model~\eqref{eq:Ham0} can be straightforwardly diagonalized 
by first mapping it to a free-fermionic model via a Jordan-Wigner transformation,
and then going in the Fourier space [a Bogoliubov rotation is also required
for the Hamiltonian~\eqref{eq:hamiltonian} with $\lambda \neq 0$].
The energy dispersion relation for the system quasiparticles is given by
$\epsilon_k  = 2 J_0 \cos (k + \phi/L)$, 
from which one can calculate the ground-state current $\langle {\cal I} \rangle = - \partial E / \partial \phi$,
with $E$ being the ground state energy.
A plot of such current, as well as of the bipartite entanglement entropy, as a function of the flux $\phi$
for a ring made of $L=4$ spins, is provided in Fig.~\ref{fig:ground_state_current}. 
As pointed out in Sec.~\ref{sec:flux_based_currents} in the dissipative case,
we observe that even in this case the current is not an entanglement related phenomenon,
in the sense that, while the current changes with $\phi$, the von Neumann entropy of the two-site reduced
density matrix is insensitive to it.
We stress that, since the gauge field enters in the model as $\phi/L$, the observed current
vanishes as $\sim 1/L$ in the macroscopic limit $L \to \infty$ (it is also possible to exploit
the gauge freedom of the model, in order to put the phase $\phi$ only in one link,
thus acting as a twist in the boundary condition of the ring).

\section{Cluster mean-field ansatz}
\label{app:mean field}

Here we report some technical details on the cluster mean-field ansatz employed in Sec.~\ref{sec:mean_field},
in order to find analytic results. Since the Hamiltonian of our minimal model contains just an onsite term
(i.e., the chemical potential term), the CMF decoupling will only affect the dissipative part of the dynamics.
Note that a single-site mean-field ansatz is not able to describe any form of directionality
and of current, since at least two sites are involved.
Our approach thus represents the minimal ansatz necessary to unveil persistent currents.
Let us thus build up a cluster of two sites $\{j,j+1\}$, such that we can write the system's
density matrix in a cluster-factorized form:
\begin{equation}
  \rho_{\rm CMF} = \bigotimes_{\cal C} \rho_{\cal C} = \bigotimes_{j\ {\rm odd}} \rho_{j,j+1} ,
\end{equation}
while interactions between each term and sites $j-1$, $j+2$ are treated as mean-field variables.
The resulting equation for the steady state can be solved in a self-consistent manner,
eliminating the mean-field variable~\cite{Tomadin_2010, Jin_2016}.

As stated in the main text, calculations are greatly simplified by the symmetries of the system,
leading to a so-called ``X-structure'' for the two-site density matrix~\cite{Amico_2004}).
In the end we find the steady state solution
\begin{equation}
  \rho_{\rm SS} = \frac{1}{Z}
  \begin{bmatrix} \;\;
	\rho_{11} & 0 & 0 & \rho_{14}\\
	0 & \rho_{22} & \rho_{23} & 0\\
	0 & \rho_{23} & \rho_{22} & 0\\
	\rho^{*}_{14} & 0 & 0 & \rho_{44} \;\;
	\end{bmatrix} \,,
\end{equation}
with
\begin{eqnarray}
  Z & = & (\alpha^2-\delta^2)^2 (\alpha^4+\delta^4+3\alpha^2\delta^2) \! + \! \mu^2(\alpha^2+\delta^2)^2, \quad \\
  \rho_{11} & = & \tfrac{1}{4}\delta^2 \big[(4\delta^2+\alpha^2)(\alpha^2-\delta^2)^2+4\delta^2\mu^2 \big], \\
  \rho_{22} & = & \tfrac{1}{4}\alpha^2 \big[ \alpha^2\delta^2(5(\alpha^2-\delta^2)+4\mu^2) \big], \\
  \rho_{44} & = & \tfrac{1}{4}\alpha^2 \big[ (4\alpha^2+\delta^2)(\alpha^2-\delta^2)^2+4\alpha^2\mu^2 \big], \\
  \rho_{23} & = & \alpha^2\delta^2(\alpha^2-\delta^2)^2, \\
  \rho_{14} & = & \tfrac{1}{2}(\alpha^2+\delta^2+i\mu)(\delta^2-\alpha^2)\alpha\delta.
\end{eqnarray}
From this solution it is possible to evaluate all the relevant quantities,
in particular one can retrieve the current in Eq.~\eqref{eq:Curr_CMF} in the main text.


\begin{thebibliography}{100}

\bibitem{Yang_1967}
  C. N. Yang, Phys. Rev. Lett. {\bf 19}, 1312 (1967).

\bibitem{Tinkham_1996}
  M. Tinkham, {\it Introduction to Superconductivity}, (McGraw-Hill, New York, 1996).

\bibitem{Imry_1997}
  Y. Imry, {\it Introduction to mesoscopic physics}, (Oxford University Press, Oxford, 1997).

\bibitem{Phillips}
  A. Ramanathan, K. C. Wright, S. R. Muniz, M. Zelan, W. T. Hill III, C. J. Lobb, K. Helmerson,
  W. D. Phillips, and G. K. Campbell,  Phys. Rev. Lett. {\bf 106}, 130401 (2011);
  K. C. Wright, R. B. Blakestad, C. J. Lobb, W. D. Phillips, and G. K. Campbell,
  Phys. Rev. Lett. {\bf 110}, 025302 (2013).

\bibitem{Dalibard_2011}
  J. Dalibard, F. Gerbier, G. Juzeli\=unas, and P. \"Ohberg,  Rev. Mod. Phys. {\bf 85}, 1523 (2011).

\bibitem{Spielman_2014}
  N. Goldman, G. Juzeli\=unas, P. \"Ohberg, and I. B. Spielman,  Rep. Prog. Phys. {\bf 77}, 126401 (2014).

\bibitem{Amico_2005}
  L. Amico, A. Osterloh, and F. Cataliotti, Phys. Rev. Lett. {\bf 95}, 063201 (2005).

\bibitem{Seaman_2007}
  B. T. Seaman, M. Krämer, D. Z. Anderson, and M. J. Holland, Phys. Rev. A {\bf 75}, 023615 (2007).

\bibitem{Amico_2014}
  L. Amico, D. Aghamalyan, F. Auksztol, H. Crepaz, R. Dumke, and L.-C. Kwek, Scientific Reports {\bf 4}, 4298 (2014).

\bibitem{Aghamalyan_2015}
  D. Aghamalyan, M. Cominotti, M. Rizzi, D. Rossini, F. Hekking, A. Minguzzi, L.-C. Kwek, and L. Amico, New J. Phys. {\bf 17}, 045023 (2015).

\bibitem{Dumke_2016}
  R. Dumke {\it et al.}, J. Opt. {\bf 18}, 093001 (2016).
  
\bibitem{Haug_2018}
  T. Haug, L. Amico, R. Dumke, and L.-C. Kwek, Quantum Sci. Technol. {\bf 3}, 035006 (2018).

\bibitem{Kasprak_2006}
  J. Kasprzak {\it et al.},  Nature {\bf 443}, 409 (2006).

\bibitem{Esslinger_2010}
  K. Baumann, C. Guerlin, F. Brennecke, and T. Esslinger,  Nature {\bf 464}, 1301 (2010).

\bibitem{Carusotto_2013}
  I. Carusotto and C. Ciuti,  Rev. Mod. Phys. {\bf 85}, 299 (2013).
  
\bibitem{Fitzpatrick_2017}
  M. Fitzpatrick, N. M. Sundaresan, A. C. Y. Li, J. Koch, and A. A. Houck,  Phys. Rev. X {\bf 7}, 011016 (2017).

\bibitem{Hartmann_2008}
  M. J. Hartmann, F. G. S. L. Brand\~a o, and M. B. Plenio, Laser Photon. Rev. {\bf 2}, 527 (2008).
  
\bibitem{TomadinRev_2010}
  A. Tomadin and R. Fazio, J. Opt. Soc. Am. {\bf 27}, A130 (2010).
  
\bibitem{Houck_2012}
  A. A. Houck, H. E. Türeci, and J. Koch,  Nat. Phys. {\bf 8}, 292 (2012).

\bibitem{Noh_2016}
  C. Noh and D. G. Angelakis,  Rep. Prog. Phys. {\bf 80}, 016401 (2016).

\bibitem{Zhou_2008}
  {L. Zhou, Z. R. Gong, Y. Liu, C. P. Sun, and F. Nori,  Phys. Rev. Lett. {\bf 101}, 100501 (2008).}
 
\bibitem{Marquardt_2013}
  M. Ludwig and F. Marquardt,  Phys. Rev. Lett. {\bf 111}, 073603 (2013).

\bibitem{Zhu_2013}
  {C. Zhu, S. Endo, P. Naidon, and P. Zhang,  Few-Body Syst. {\bf 54}, 1921 (2013).}

\bibitem{Saffmann_2010}
  M. Saffman, T. G. Walker, and K. Mølmer,  Rev. Mod. Phys. {\bf 82}, 2313 (2010).

  \bibitem{Muller_2012}
  M. M\"uller, S. Diehl, G. Pupillo, and P. Zoller,  Adv. At. Mol. Opt. Phys. {\bf 61}, 1 (2012).
  
\bibitem{Diehl_2008}
  S. Diehl, A. Micheli, A. Kantian, B. Kraus, H. P. B\"uchler, and P. Zoller,  Nat. Phys. {\bf 4}, 878 (2008).

\bibitem{Verstraete_2009}
  F. Verstraete, M. M. Wolf, and J. I. Cirac,  Nat. Phys. {\bf 5}, 633 (2009).

\bibitem{Diehl_2011}
  S. Diehl, E. Rico, M. A. Baranov, and P. Zoller,  Nat. Phys. {\bf 7}, 971 (2011).

\bibitem{Bardyn_2013}
  C.-E. Bardyn, M. A. Baranov, C. V. Kraus, E. Rico, A. Imamoglu, P. Zoller, and S. Diehl,
  New J. Phys. {\bf 15}, 085001 (2013).
  
\bibitem{Iemini_2015}
  F. Iemini, D. Rossini, R. Fazio, S. Diehl, and L. Mazza, Phys. Rev. B {\bf 93}, 115113 (2016).
  
\bibitem{Ozawa_2016}
  T. Ozawa, H. M. Price, N. Goldman, O. Zilberberg, and I. Carusotto,
  Phys. Rev. A {\bf 93}, 043827 (2016)

\bibitem{Pichler_2015}
  H. Pichler, T. Ramos, A. J. Daley, and P. Zoller, Phys. Rev. A {\bf 91}, 042116 (2015).

\bibitem{Lodahl_2017}
  P. Lodahl, S. Mahmoodian, S. Stobbe, A. Rauschenbeutel, P. Schneeweiss, J. Vold, H. Pichler, and P. Zoller,
  Nature {\bf 541}, 473 (2017).

\bibitem{Vermersch_2017}
  B. Vermersch, P.-O. Guimond, H. Pichler, and P. Zoller,  Phys. Rev. Lett. {\bf 118}, 133601 (2017).
  
\bibitem{Metelmann_2015}
  A. Metelmann and A. A. Clerk, Phys. Rev. X {\bf 5}, 021025 (2015).

\bibitem{Petruccione_2002}
  H.-P. Breuer and F. Petruccione, {\it The Theory of Open Quantum Systems}
  (Oxford University Press, New York, 2002).

\bibitem{Nielsen-Chuang}
  M. A. Nielsen and I. L. Chuang, {\it Quantum computation and quantum information: 10th Anniversary Edition},
  Cambridge University Press, NY, USA (2010).
  
\bibitem{Peres-Horodecki}
  A. Peres,  Phys. Rev. Lett. {\bf 77}, 1413 (1996);
  M. Horodecki, P. Horodecki, and R. Horodecki, Phys. Lett. A {\bf 223}, 1 (1996).

\bibitem{Tomadin_2010}
  A. Tomadin, V. Giovannetti, R. Fazio, D. Gerace, I. Carusotto, H. E. Tu\"reci, and A. Imamoglu,
  Phys. Rev. A {\bf 81}, 061801 (2010).

\bibitem{Jin_2016}
  J. Jin, A. Biella, O. Viyuela, L. Mazza, J. Keeling, R. Fazio, and D. Rossini,
  Phys. Rev. X {\bf 6}, 031011 (2016).

\bibitem{Amico_2004}
  L. Amico, A. Osterloh, F. Plastina, R. Fazio, G. M. Palma,
  Phys. Rev. A. {\bf 69}, 022304 (2004).

\bibitem{MPS_dissipative}
  F. Verstraete, J. J. Garcia-Ripoll, and J. I. Cirac, Phys. Rev. Lett. {\bf 93}, 207204 (2004);
  M. Zwolak and G. Vidal, Phys. Rev. Lett. {\bf 93}, 207205 (2004).
  
\bibitem{Bloch_2013}
  M. Aidelsburger, M. Atala, M. Lohse, J. T. Barreiro, B. Paredes, and I. Bloch,
  Phys. Rev. Lett. {\bf 111}, 185301 (2013).

\bibitem{Ketterle_2013}
  H. Miyake, G. A. Siviloglou, C. J. Kennedy, W. C. Burton, and W. Ketterle,
  Phys. Rev. Lett. {\bf 111}, 185302 (2013).

\bibitem{Fallani_2015}
  M. Mancini, G. Pagano, G. Cappellini, L. Livi, M. Rider, J. Catani, C. Sias, P. Zoller,
  M. Inguscio, M. Dalmonte, and L. Fallani, Science {\bf 349}, 1510 (2015).

\bibitem{Spielman_2015}
  B. K. Stuhl, H.-I. Lu, L. M. Aycock, D. Genkina, and I. B. Spielman, Science {\bf 349}, 1514 (2015).

\bibitem{Peano_2015}
  V. Peano, C. Brendel, M.Schmidt, and F. Marquardt,
  Phys. Rev. X {\bf 5}, 031011 (2015).

\bibitem{Barik_2018}
  S. Barik, A. Karasahin, C. Flower, T. Cai, H. Miyake, W. DeGottardi, M. Hafezi, and E. Waks,
  Science {\bf 369}, 666 (2018).

\bibitem{Ozawa_2018}
  T. Ozawa, H. Price, A. Amo, N. Goldman, M. Hafezi, L. Lu, M. Rechtsman, D. Schuster, J. Simon, O. Zilberberg, and I. Carusotto,
  arXiv:1802.04173
  
\end{thebibliography}
\end{document}